\documentclass[prd,article,11pt]{revtex4}%
\usepackage{amsmath, amsthm, amsfonts}
\usepackage[spanish,activeacute,english]{babel}
\usepackage{amsmath}
\usepackage{amsfonts}
\usepackage{amssymb}
\usepackage{graphicx}
\usepackage{cancel}
\usepackage{appendix}
\usepackage{graphicx}
\usepackage{vmargin}
\usepackage{makeidx}
\usepackage{float}
\usepackage{subfig}
\usepackage{braket}
\usepackage{dsfont}
\usepackage{hyperref}
\usepackage{color}
\usepackage{amsmath}
\usepackage{graphicx}
\usepackage{epsfig}
\usepackage{dcolumn}
\usepackage{bm}
\usepackage{dcolumn}
\usepackage{amsfonts}
\usepackage{amssymb}%
\providecommand{\U}[1]{\protect\rule{.1in}{.1in}}
\providecommand{\U}[1]{\protect\rule{.1in}{.1in}}

\theoremstyle{definition}

\theoremstyle{remark}

\begin{document}
\title{Modelos SIR modificados para la evoluci\'{o}n del COVID19 }
\author{Nana Cabo Bizet \footnote{nana@fisica.ugto.mx} y Alejandro Cabo Montes de
Oca\footnote{cabo@icimaf.cu} \medskip}
\affiliation{$^{1}$ \textit{Departamento de F\'{\i}sica, DCI, Campus Le\'on, Universidad de
Guanjuato, \ CP. 37150, Le\'on, Guanajuato, M\'exico }}
\affiliation{$^{2}$\textit{Departamento de F\'isica Te\'orica, Instituto de
Cibern\'{e}tica, Matem\'{a}tica y F\'{\i}sica, Calle E, No. 309, Vedado, La
Habana, Cuba. }}

\begin{abstract}
Estudiamos el modelo epidemiol\'{o}gico SIR, con tasa de contagio variable,
aplicado a la evoluci\'{o}n del COVID 19 en Cuba.  Se resalta que un aumento de car\'{a}cter 
predictivo de este modelo depende de la comprensi\'{o}n de la din\'{a}mica que determina la evoluci\'{o}n
temporal de la tasa de contagio $\beta^{*}$. Se formula un modelo
semi-emp\'{\i}rico para dicha din\'{a}mica; donde alcanzar $\beta^{*}\approx0$
debido al aislamiento se logra despu\'{e}s del tiempo de duraci\'{o}n media de
la enfermedad $\tau=1/\gamma$, en que ha decrecido el n\'{u}mero de infectados
en las familias confinadas. Se considera que $\beta^*(t)$ debe tener una
disminuci\'{o}n brusca en el d\'{\i}a de inicio del confinamiento y reducirse
hasta anularse al final del intervalo $\tau$. El an\'alisis logra describir
apropiadamente la curva de infecci\'{o}n de Alemania. El mismo modelo se
aplica a predecir una curva de infecci\'{o}n para Cuba,  el cual estima un m\'aximo de
infectados que depende de cuan riguroso resulta el aislamiento y que su valor
podr\'ia ir de  1,000 a 2,000 casos si este ocurre a mediados de mayo 2020.
Esto es  sugerido por la curva de la raz\'on entre
los casos detectados diariamente y el total. Se considera la raz\'{o}n
$k$ entre el n\'{u}mero de infectados observados y el n\'umero total de
infectados menor que la unidad. El
reducido valor de $k$ disminuye los valores m\'{a}ximos obtenidos cuando
$\beta^{*}-\gamma>0$. En la regi\'on lineal con $k$ constante
la evoluci\'on de los infectados y recuperados observados es independiente de $k$. Se estudia tambi\'en el valor de $\beta^*$ a pedazos, por
intervalos de tiempo, ajustando a los datos de Cuba, Alemania y Corea del Sur.
En este esquema se compara la extrapolaci\'on de la evoluci\'on de Cuba con
tasa de contagio al d\'{\i}a 16.04.20 con la obtenida suponiendo a finales de
abril una cuarentena estricta. La evoluci\'on de este SIR con tasa variable
describe correctamente las curvas de evoluci\'on de infectados. Enfatizamos que el m\'aximo deseado de la curva de infectados del SIR no
es el m\'aximo est\'andard con $\beta^{*}$ constante, sino uno logrado debido
a la cuarentena cuando localmente se logra $\tilde R_{0}=\beta^{*}/\gamma<1$.
Para todos los pa\'{\i}ses que han salido de la epidemia estos m\'aximos se
encuentran en la regi\'on en que las ecuaciones SIR son lineales.


\end{abstract}
\maketitle

\newpage

\section{ Introducci\'{o}n}

La pandemia asociada la COVID 19 est\'a siendo investigada con gran intensidad
hoy d\'ia y la descripci\'on de sus singulares propiedades ampliamente
difundidas \cite{1,2,3,4,5,6,7,8,9,10,11,12,13,14,15,16,17,18,19,20,21,22,23}. El presente trabajo est\'{a} dedicado
a explorar aspectos de esta epidemia. Consideramos un modelo determinista
empleado en el estudio de epidemias, el SIR (siglas de Susceptibles,
Infectados y Recuperados). Los par\'{a}metros que controlan dichos modelos son
$\beta$ y $\gamma$. El primero define el n\'{u}mero de personas susceptibles
por unidad de tiempo que enferman por unidad de persona susceptible y por
unidad de persona infectada. Este tambi\'en se utiliza despu\'es de multiplicado  por el n\'umero
de la poblaci\'on, en las unidades $\beta^{*}=\beta N$. El segundo determina
el n\'{u}mero de personas por unidad de tiempo que se recuperan entre el
n\'{u}mero de personas infectadas. Estimamos que la aplicaci\'{o}n del SIR en
una efectiva predicci\'{o}n de los datos necesita la clarificaci\'on un
aspecto central: el como estimar la dependencia temporal de los par\'ametros
$\beta$ y $\gamma$. En particular aqu\'{\i} nos enfocamos en el estudio de la
evoluci\'on del primero de estos.

En la secci\'{o}n I se exponen las ecuaciones del modelo SIR y se presentan
sus constantes para el caso de Cuba. Las soluciones obtenidas se ajustan a los
datos de n\'{u}mero de infectados activos que se reportan diariamente.
Destacamos las condiciones en que se obtiene un m\'{a}ximo valor del
n\'{u}mero de infectados en circunstancias realistas. Dichas
caracter\'{\i}sticas son conocidas, pero creemos \'{u}til el insistir sobre
ellas, se destaca que siempre que las ecuaciones indican infecci\'{o}n, se
cumple $\beta^{*}>\gamma$. En esa situaci\'on el m\'{a}ximo del n\'umero de
infectados siempre alcanza un valor del orden de la poblaci\'{o}n $N$, lo que
para Cuba resultar\'{\i}a en millones. Resaltamos que esos m\'{a}ximos, no son
los que se reportan en los pa\'{\i}ses que actualmente rebasan la epidemia
(por ej. China, Corea del Sur y Alemania), los cuales presentan m\'{a}ximos
del n\'{u}mero de infectados mucho menores que la poblaci\'{o}n.

En la secci\'{o}n II se considera el mismo modelo SIR pero en el cual se
introduce una ya reconocida propiedad de la presente epidemia: la raz\'{o}n
$k$ entre el n\'{u}mero de infectados observados (por los sistemas de salud) y
el total de infectados, es un n\'{u}mero que se estima en el intervalo de
$0.1$ a $0.2$ \cite{5,12}. En esta secci\'{o}n se muestran las curvas
soluci\'{o}n despu\'{e}s de ajustar los datos para los infectados reales con
los datos observados despu\'{e}s de divididos por $k$. Los enormes n\'{u}meros
m\'{a}ximos de enfermos observados se reducen a medida que $k$ disminuye. Este
efecto es una consecuencia directa de la no linealidad del sistema de
ecuaciones. Si el sistema fuera linear el m\'{a}ximo no deber\'{\i}a cambiar,
pues al variar las condiciones iniciales en ser divididas por $k$, la
soluci\'{o}n para el n\'{u}mero total ser\'{\i}a proporcional a la anterior.
Luego, despu\'{e}s de multiplicar por $k$ el m\'{a}ximo n\'{u}mero de
infestados totales para obtener el observado, se volver\'{\i}a a obtener el
mismo valor. Sin embargo, como el sistema de ecuaciones es no lineal (el
n\'{u}mero de infectados no puede sobrepasar la poblaci\'{o}n del pa\'{\i}s)
la no linealidad resulta capaz de reducir el m\'{a}ximo. Sin embargo nuestro
inter\'{e}s son las condiciones realistas donde el m\'{a}ximo est\'{a} muy
lejos de acercarse a la poblaci\'{o}n del pa\'{\i}s, por lo tanto estamos en la regi\'on
lineal. Esto significa que nuestra predicci\'on para infectados y recuperados observados
si $k$ es constante es independiente de su valor. En la secci\'{o}n III discutimos las condiciones de aparici\'{o}n de esos
m\'{a}ximos reducidos.

La secci\'{o}n III comienza discutiendo las soluciones del sistema SIR en los
casos de que el n\'{u}mero de infectados $I$ es mucho m\'{a}s peque\~{n}o que
la poblaci\'{o}n $N$. En esta situaci\'on el n\'{u}mero de susceptibles en esa
etapa se puede aproximar por la poblaci\'{o}n del pa\'{\i}s $N$. Se presenta
entonces las conocidas soluciones expl\'{\i}citas del problema para $\beta$ y
$\gamma$ constantes. Estas son exponenciales cuyo tiempo caracter\'{\i}stico
de crecimiento o decrecimiento es determinado por el \'{u}nico n\'{u}mero
$\beta^{*}-\gamma$. Si esta cantidad es positiva la soluci\'{o}n del problema
crece indefectiblemente, independiente de los valores de $I$ y $R$ en el momento
dado. Esta es una propiedad importante de la din\'{a}mica considerada. Para
que $I$ decrezca en esta zona de bajos valores ($I \ll N$) es estrictamente
necesario que los valores locales de $\beta-\gamma$ resulten negativos o que
el par\'ametro $(\tilde R_{0}<1)$. Por tanto, en todos los pa\'{\i}ses en que
se ha logrado la recuperaci\'{o}n lo que se ha alcanzado son valores negativos
de esta cantidad. Sin embargo, aunque el pa\'{\i}s imponga aislamiento total a
partir de cierto d\'{\i}a, las curvas de infectados en ning\'{u}n caso
comienzan a bajar. Parece l\'{o}gico suponer que esas medidas deben forzar la
validez de la condici\'on $\beta=0$, o sea, ausencia de transmisibilidad. Las
causas de esta aparente contradicci\'on se analizan en la siguiente secci\'{o}n.

En esa secci\'{o}n se presenta una raz\'{o}n por la cual el valor de $\beta=0$
no es inmediatamente determinado por las medidas de aislamiento. La idea se
basa en que el tiempo en que los enfermos sufren el padecimiento se estima en
un intervalo $\tau$ de $15$ a $20$ d\'{\i}as. En este caso el enclaustramiento
de las familias, indica que en gran cantidad de ellas pueden existir enfermos
asintom\'{a}ticos. Por tanto el n\'{u}mero positivo de infectados por unidad
de tiempo no se anula al momento, si no que debe variar discontinuamente de
valor (el entorno de cada infestado cambia abruptamente). Por tanto a partir
del instante en que comienza el aislamiento y durante veinte d\'{\i}as, el
n\'{u}mero $\beta$ no debe ser nulo sino debe decrecer de cierta forma hasta
anularse a los $20$ d\'{\i}as aproximadamente. La dependencia temporal en ese
intervalo sin embargo no es conocida. Solo es de esperar que ella tenga un
salto brusco el d\'{\i}a del aislamiento debido a que las condiciones de
contacto de los infectados con su entorno cambiaron dr\'{a}sticamente. En la
secci\'{o}n se analiza adem\'{a}s la curva de infecci\'{o}n de China la que
permite estimar el valor de $\gamma$ en alrededor de $1/20$. Este modelo
cualitativo de la evoluci\'{o}n de $\beta$ bajo condiciones de aislamiento se
aplica a describir las curvas de infectados de Alemania y Cuba en la
subsecci\'{o}n A.

La subsecci\'{o}n A primeramente fija los par\'{a}metros $\beta$ y $\gamma$ en
el comienzo de crecimiento exponencial de la infecci\'{o}n en Alemania.
Posteriormente se utiliza la informaci\'{o}n de que en ese pa\'{\i}s las
medidas de aislamiento comenzaron cerca del $20$ de marzo. Utilizando que la
curva de infecci\'{o}n muestra un cambio en pendiente ese d\'{\i}a, se estima
entonces el nuevo valor de $\beta$ al que "salta" esta magnitud bruscamente al
imponer el aislamiento. Posteriormente se asume que $\beta$ disminuye
linealmente durante un tiempo de vida de la enfermedad de cerca de $20$
d\'{\i}as. Al final de ese per\'{\i}odo de confinamiento, el valor disminuye
dr\'{a}sticamente a cero. Ajustando la dependencia temporal de $\beta$ para
tiempos posteriores al instante de aislamiento, hasta su disminuci\'{o}n
brusca a cero 20 d\'{\i}as despu\'{e}s, se logra entonces describir la curva
de infecci\'{o}n de Alemania. El mismo modelo se aplica a la curva de
infecci\'{o}n de Cuba. Se toma el 24 de marzo como el d\'{\i}a de
imposici\'{o}n del aislamiento y los par\'{a}metros $\beta$ y $\gamma$ se
derivan de describir aproximadamente los datos diarios de infectados. Se
obtiene un m\'{a}ximo de infectados que depende del rigor con que se imponga
el aislamiento. Los datos sugieren que si el aislamiento resultara
gradualmente menos efectivo  se podr\'{\i}a obtener desde un m\'{a}ximo del
orden de $1200$ casos cerca del 1 de mayo a  uno de alrededor de 2000 casos
para el 12  de mayo. El an\'{a}lisis en  general  sugiere que los
m\'{a}ximos de la epidemia en Cuba  tienen buena probabilidad de suceder en ese
rango de valores.

En la \'{u}ltima subsecci\'{o}n B se estudian las curvas de infectados de
Cuba, Alemania y Corea del Sur. El m\'{e}todo que se sigue es el de explorar
la tasa de contagio variable $\beta(t)$ haciendo ajustes locales, al dividir
los datos en segmentos. Se estudian los datos de Cuba, considerando dos
escenarios: En el primero la tasa de contagio continua siendo la tasa al
d\'{\i}a de hoy (16 de Abril 2020), la epidemia se extender\'{\i}a a junio y los enfermos
detectados en el pico rondar\'{\i}an los 500,000. En el segundo escenario a
finales de abril la cuarentena estricta logra $\tilde{R}_{0}<1$ y por tanto el
pico de la epidemia se alcanza en mayo, y los enfermos detectados en el pico
rondan los varios miles (m\'{a}s de 3000). Se comprueba que la cuarentena en
el caso de Alemania y Corea del Sur ya alcanz\'{o} la regi\'{o}n $\tilde
{R}_{0}<1$ ($\beta^{\ast}-\gamma<0$), que es lo deseado. Estos \'{u}ltimos
resultados enfatizan la importancia de una cuarentena estricta. Todo lo
anterior se resume en nuestra conclusiones en la \'{u}ltima secci\'{o}n.

\section{El modelo SIR}

El modelo``Suceptible Infected Recovered" (SIR) es uno de los m\'{a}s
sencillos y claves en los estudios epidemiol\'{o}gicos para la propagaci\'{o}n
de enfermedades \cite{1}. En el mismo la poblaci\'{o}n total $N$ se divide en
tres grupos: suceptibles ${S}$, infectados ${I}$ y recuperados ${R}$.
Denominemos por ${N}$ la poblacion del pa\'{\i}s considerado. Las ecuaciones
diferenciales del modelo que gobiernan la evoluci\'{o}n de la epidemia
est\'{a}n dadas por%
\begin{align}
\frac{\partial S}{\partial t}  &  = -\beta S I,\label{sir1}\\
\frac{\partial I}{\partial t}  &  = \beta\ S I-\gamma I,\label{sir2}\\
\frac{\partial R}{\partial t}  &  = \gamma I. \label{sir3}%
\end{align}

Las unidades de esta ecuaci\'{o}n corresponden a $personas(P)$ para las
magnitudes de grupos poblacionales son $[S]=[I]=[R]=P$. La unidad de $tiempo$
$(T)$ es un d\'{\i}a y $\beta$ tiene unidades: $[\beta]=1/(PT)$. Las unidades
de $\gamma$ son $[\gamma]=1/T$. El n\'{u}mero $\gamma$ representa la
proporci\'{o}n de recuperados por unidad de tiempo entre la poblaci\'{o}n
infectada. Consideraremos la cantidad $\gamma$ estimada de las curvas de
recuperacion en pa\'{\i}ses que han sobrepasado ya la epidemia, como es el
caso de China, el resultado haciendo un ajuste de m\'{\i}nimos cuadrados es
$\gamma=1/20$.

Estas ecuaciones tambi\'{e}n pueden darse en unidades donde las poblaciones
sean adimensionales $\tilde{S}=S/N,\tilde{I}=I/N,\tilde{R}=R/N$:
\[
\frac{\partial\tilde{S}}{\partial t}=-\frac{\tilde{\beta}\tilde{S}\tilde{I}%
}{N},\,\,\,\,\frac{\partial\tilde{I}}{\partial t}=\frac{\tilde{\beta}\tilde
{S}\tilde{I}}{N}-\gamma\tilde{I},\,\,\,\,\,\frac{\partial\tilde{R}}{\partial
t}=\gamma\tilde{I}.
\]
Las unidades de $\tilde{\beta}=\beta N^{2}=\beta^{\ast}N$ est\'{a}n dadas por
$[\tilde{\beta}]=P/T$. Tambi\'{e}n emplearemos $\beta^{\ast}=\beta\ast N$ con
unidades $[\beta^{\ast}]=1/T$.

Algo a destacar es que las magnitudes $S,I$ y $R$ en las ecuaciones
(\ref{sir1},\ref{sir2},\ref{sir3}) son los n\'umeros de personas susceptibles,
infectadas y recuperadas que existen realmente en toda la poblaci\'on. Sin
embargo, los datos de que se disponen para resolver las ecuaciones a partir de
sus condiciones iniciales, en muchos casos son solo la poblaci\'on total del
pa\'{\i}s y los n\'{u}meros de infectados y recuperados que detecta el sistema
de Salud. En esta secci\'on consideraremos que de los datos observados, el
\'unico que es exacto es la poblaci\'on del pa\'{\i}s que define a $S$ al
inicio. Se toma en cuenta una poblaci\'{o}n de 12 millones de personas para
Cuba y de 83 millones para Alemania.

A medida de comenzar con un modelo de juguete, consideremos los datos
presentados a partir del 16.03.20 al d\'{\i}a 16.04.2020, calculamos los
siguientes par\'ametros globales del modelo SIR de Cuba:
\begin{equation}
\tilde{R}_{0}=4.04413,\,\,\beta^{*}=0.202206\, \frac{1}{T}, \, \, \gamma
=\frac{1}{20} \, \frac{1}{T}.
\end{equation}
\begin{figure}[h]
\begin{center}
\includegraphics[width=.5\textwidth]{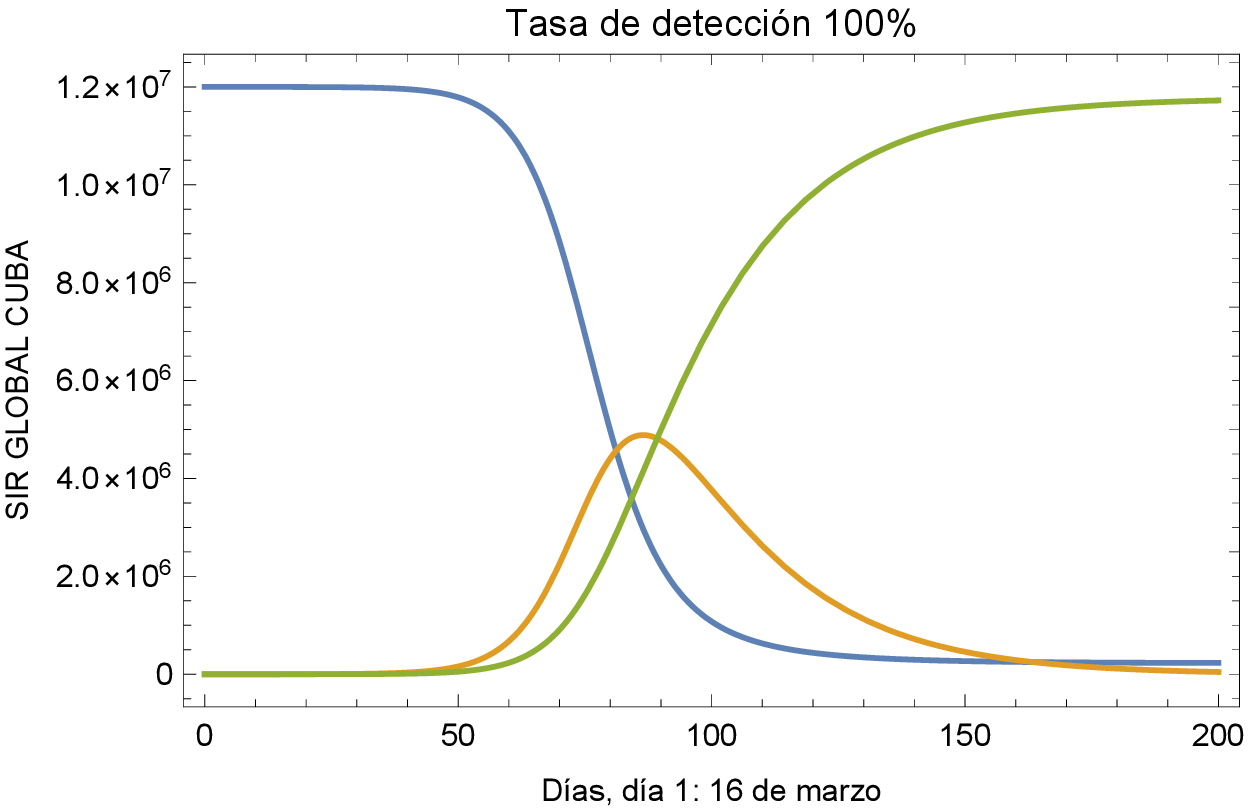}
\includegraphics[width=.5\textwidth]{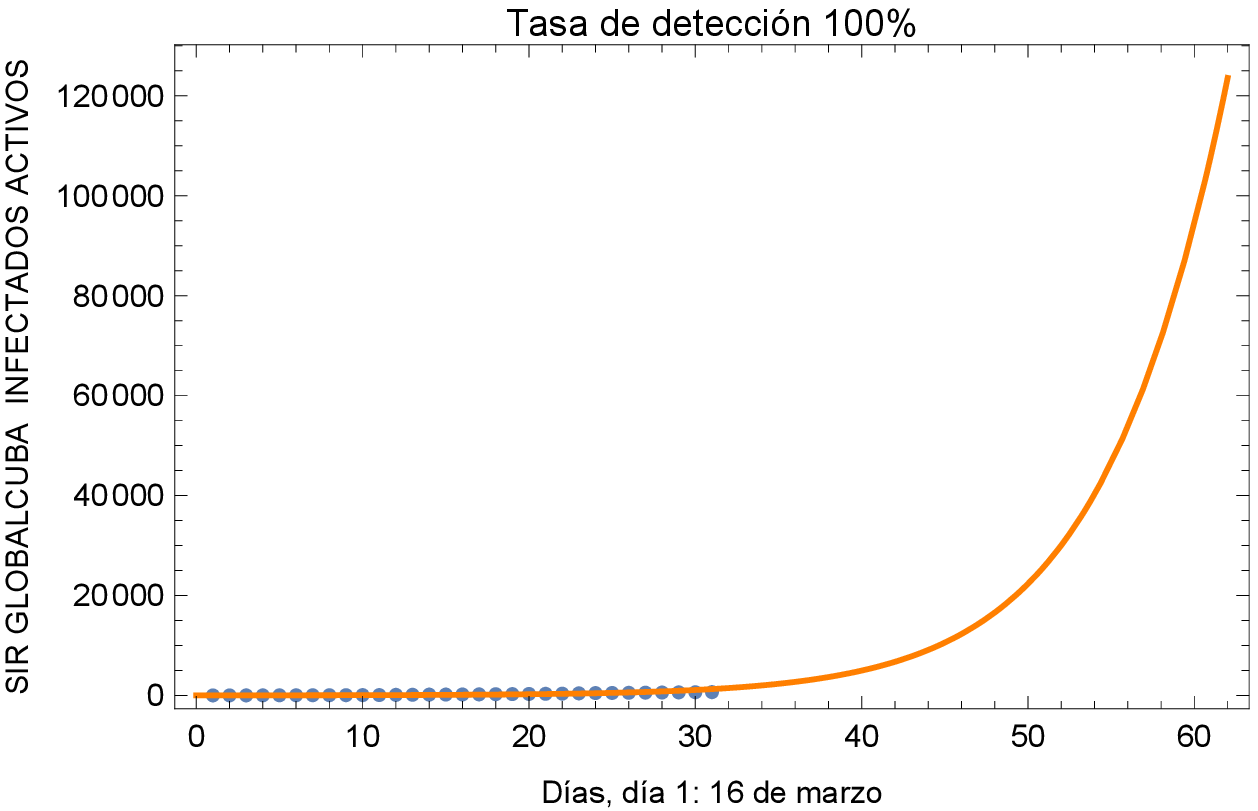}
\end{center}
\caption{ {\protect\footnotesize Curvas \textquotedblleft ideales" de la
evoluci\'{o}n del SIR en Cuba, considerando tasa de contagio constante al
d\'{\i}a de hoy, y tasa de detecci\'{o}n del 100\%. \ \ \ En el panel de arriba, \ la curva azul corresponde a los susceptibles, la naranja a los
recuperados y la amarilla \ a los infectados activos. Los n\'{u}meros
obtenidos son gigantescos, porque no se considera el efecto de la cuarentena
en la disminuci\'{o}n de $\beta^{\ast}$, ni el hecho de que s\'{o}lo se
detectan a nivel mundial del orden del 10\% de los casos. El panel de abajo muestra un intervalo de tiempo m\'as peque\~no. }}%
\label{figu1}%
\end{figure}

Esta dependencia se obtiene ajustando a los datos experimentales, es decir la
dependencia de los infectados activos en t\'erminos del tiempo al comienzo de
la epidemia. Se considera una tasa de detecci\'on del 100 porciento. Esto es
una valoraci\'on muy primitiva de la pandemia, que debe corregirse tomando una
tasa de detecci\'on del orden del 10\% \cite{12} y en nuestra opini\'on
considerando valores variables de $\beta$ para reflejar las medidas de
contenci\'on. Los resultados se muestran en la figura \ref{figu1}.



\section{Modelos SIR y raz\'on entre infectados detectados y totales}

Consideremos ahora que el n\'{u}mero total de personas infectadas $I(t)$ se
desconoce. Esto sucede debido a que hay personas que enferman y sanan sin ser
reportadas, debido a presentar s\'{\i}ntomas leves, denotemos por $I_{o}(t)$
al n\'{u}mero de personas infectadas observado. La relaci\'{o}n $k$ entre
$I_{0}(t)$ e $I(t)$ asumiremos que es una constante estimada en la literatura
\cite{5,12} por lo cual
\[
I_{o}(t)=k\text{ }I(t),\,\,\,\,\,k=\frac{r}{1+r},
\]
donde $r$ es la raz\'{o}n entre el n\'{u}mero de infectados observados y el
n\'{u}mero de los \ no observados, la cual ha sido estimada en la referencia
\cite{5} a un valor en el rango $(0.1,0.2)$. Los casos observados est\'{a}n
entre un d\'{e}cimo y dos d\'{e}cimos de los casos no observados. Adoptaremos
un valor de $k=0.2$ cercano a $r=1/5$, el cual es simplemente un estimado. El
cociente entre el n\'{u}mero de recuperados observados a un tiempo dado y el
n\'{u}mero total de recuperados $k^{\ast}$ podr\'{\i}a ser tambi\'{e}n una
constante en la zona de tiempos peque\~{n}os, ya que la soluci\'{o}n es
exponencial. No asumiremos que $k^{\ast}$ coincide con el valor de $k$, aunque
en los gr\'{a}ficos representaremos tambi\'{e}n la magnitud
\[
R_{o}(t)=k\text{ }R(t),
\]
que constituyen los recuperados observados en el caso de que $k^{\ast}=k$. Por
lo cual este n\'{u}mero no constituye una predicci\'{o}n para el n\'{u}mero de
recuperados observados.

Consideremos la soluci\'{o}n del sistema de ecuaciones (\ref{sir1}%
,\ref{sir2},\ref{sir3}) que describa aproximadamente la lista de valores para
el n\'{u}mero de los infectados activos y sus incrementos diarios observados
por el Sistema de Salud de Cuba entre los d\'{\i}as 11.03.20 y 3.04.20. La
soluci\'{o}n del sistema debe describir los valores de dicha lista despu\'{e}s
de divididos por el factor $k=0.2$
\begin{equation}
I(t)=\frac{1}{k}I_{o}(t).
\end{equation}
Los datos para el n\'{u}mero de infectados crecen r\'{a}pidamente en forma
compatible con la conocida evoluci\'{o}n exponencial de las soluciones SIR
cuando la cantidad de infectados es peque\~{n}a. El valor de $\gamma$ describe
un decaimiento exponencial de los infectados si no hay contagio $(\beta=0).$
Dado que el tiempo t\'{\i}pico en que se cura cada enfermo es de alrededor de
15 a 20 d\'{\i}as, en esta secci\'{o}n adoptaremos el estimado de
$\gamma=\frac{1}{15}=0.066.$ \ Se obtuvieron valores de $\beta$ que permiten
una aproximaci\'{o}n de los datos observados para $I_{o}(t)$.

\begin{figure}[h]
\begin{center}
\includegraphics[width=.6\textwidth]{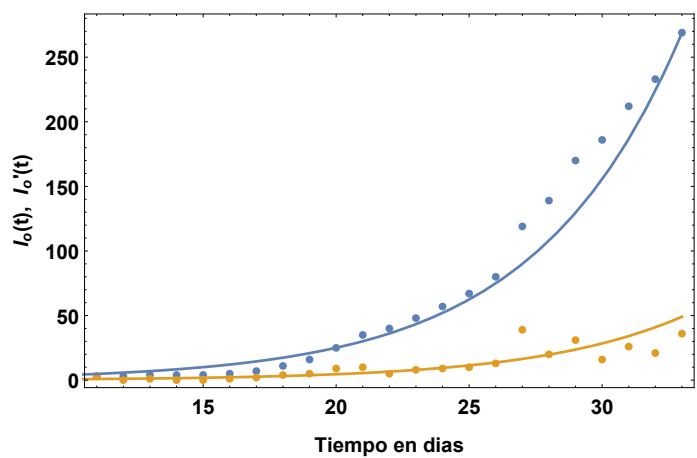}
\end{center}
\caption{{\protect\footnotesize La figura muestra la soluci\'{o}n del sistema
SIR al determinar las condiciones iniciales que ajustan la soluci\'{o}n a los
valores observados de la curva de infectados en Cuba \ (www.worldometer.info,
\cite{11}).  La curva y puntos  azules corresponden  a los infectados
predichos y observados, respectivamente, y la curva y puntos  amarillos a la
derivada respecto al tiempo predichos y observados.  Las predicciones de la
curva continua para tiempos mayores a los usados para este gr\'{a}fico, al ser
comparados con los datos que van apareciendo cada d\'{\i}a, brindan una idea
acerca del funcionamiento de las medidas de confinamiento a partir del
d\'{\i}a 24 de Marzo del 2020. }}%
\label{fig3}%
\end{figure}

Se resolvi\'{o} entonces el sistema de ecuaciones SIR fijando las condiciones
iniciales para $I(t)$ en el tiempo nulo $I(0)$. Estas condiciones iniciales se
ajustaron con vistas a reproducir los datos reportados. Los valores obtenidos
de los par\'{a}metros fueron
\[
\beta^{\ast}=a+\gamma=0.212073\,,\gamma=0.066,\,\,\beta=\frac{\text{ }%
\beta^{\ast}}{\text{ }N}=1.76727\times10^{-8}.
\]
La soluci\'{o}n obtenida para $I(t)$ y su derivada se muestran en la figura
\ref{fig3}. 

\begin{figure}[h]
\begin{center}
\includegraphics[width=.6\textwidth]{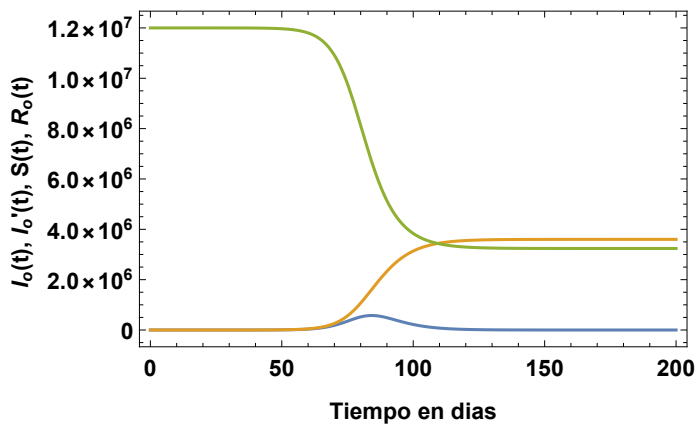}
\end{center}
\caption{ {\protect\footnotesize Se plotean $I_{o}(t),$ $S(t),$ y $R_{o}(t)$
en un intervalo de 200 d\'{\i}as. La curva azul corresponde a los infectados,
la naranja a los susceptibles y la amarilla a los recuperados. Puede
observarse que el n\'{u}mero de infectados observados tiene un m\'{a}ximo del
orden de varios cientos de miles. Cabe subrayar que estos altos n\'{u}meros
ocurren siempre que los valores de $\beta^{\ast}$ y $\gamma$ sean constantes y
exista infecci\'{o}n creciente: $\beta^{\ast}>\gamma$. Para detener la
infecci\'{o}n a bajos valores de infectados es necesario lograr $\beta^{\ast}$
se haga menor que $\gamma$. }}%
\label{fig5}%
\end{figure}

Puede observarse que la evoluci\'{o}n presenta el car\'{a}cter exponencial en
la regi\'{o}n de tiempos anterior al confinamiento, pese a las fluctuaciones
de los datos. Para un intervalo de tiempo mayor: $(0,200),$ la evoluci\'{o}n
temporal para el n\'{u}mero de infectados muestra un m\'{a}ximo como se
ilustra en la figura \ref{fig5}, que presenta  adem\'{a}s las curvas del
n\'{u}mero de susceptibles y recuperados (considerando $k^{\ast}=k$).

\begin{figure}[h]
\begin{center}
\includegraphics[width=.75\textwidth]{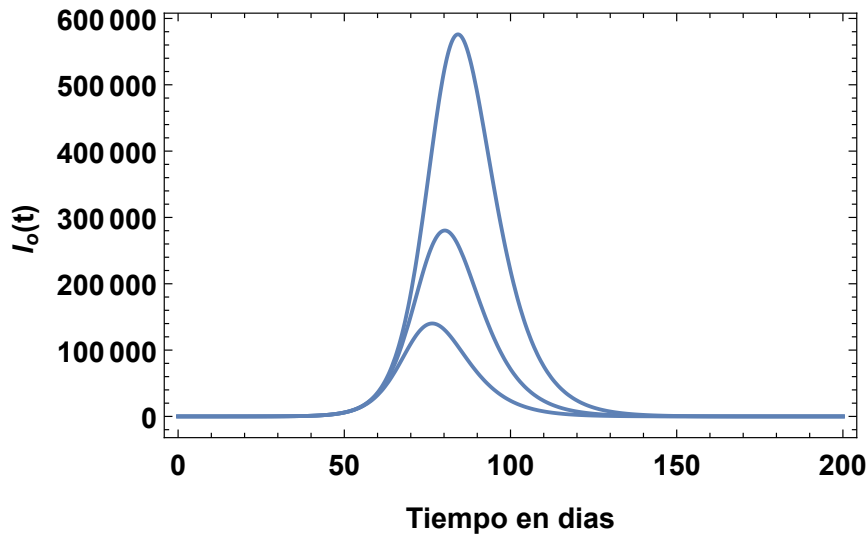}
\end{center}
\caption{{\protect\footnotesize  La figura muestra las curvas de infectados
asociadas a tres valores distintos de la relaci\'{o}n entre el n\'{u}mero de
infectados observados y la total: $k=0.411,0.2,0.1$ que corresponden en orden
a las curvas con valores crecientes de su m\'{a}ximo. Debido a la no
linealidad del sistema de ecuaciones SIR, la cantidad m\'{a}xima de infectados
decrece con la disminuci\'{o}n del n\'{u}mero $k$ de infectados no observados
por cada infectado total. La no linealidad del sistema se sigue de que el
n\'{u}mero de enfermos no puede superar a la poblaci\'{o}n. }}%
\label{fig6}%
\end{figure}

Las curvas ploteadas en un intervalo de $200$ d\'ias, muestran los resultados
para el n\'umero de susceptibles, infectados total $S$ y el n\'umero de
infectados observado $I_{o}$ que efectivamente son reportados. Finalmente,
como antes se indic\'o, $R_{o}(t)=k$ $R(t)$ representa el n\'umero de
recuperados, asumiendo que $k^{\ast}=k$. La curvas en figura \ref{fig6}
muestran $I_{o}(t)$ para tres valores distintos de $k$. Se puede notar que
bajo las suposiciones hechas, los picos del n\'umero de enfermos se
pronostican a ocurrir entre $80$ a $90$ d\'{\i}as a partir del $11$ de Marzo y
corresponden a cerca de cientos de miles de personas, en el caso de que no se
tomen medidas que puedan lograr $\beta^{*}-\gamma<0$.

Los valores de los par\'{a}metros fueron $k=0.411,0.2,0.1$, fijando $\beta$ y
$\gamma$. La curva con valor m\'{a}s bajo del m\'{a}ximo se corresponde a
$k=0.1$, la asociada al m\'{a}ximo de valor intermedio $k=0.2$ y la que posee
el m\'{a}s alto valor a  $k=0.411$.

Quisieramos subrayar que la disminuci\'on del valor del m\'aximo con el
aumento de $1/k$, es un efecto puramente no lineal. Esto ocurre debido a que
usamos datos experimentales para los n\'umeros totales de infectados siendo
proporcionales con $1/k$. En el caso de que las ecuaciones fueran lineales, el
m\'aximo de $I_{o}(t)$ hubiera resultado id\'entico para las tres curvas. Sin
embargo, la no linealidad del sistema tiene un efecto relevante en la
reducci\'on del m\'inimo en el n\'{u}mero $1$/$k.$ La no linealidad se produce
debido a que el n\'umero total de infectados no puede nunca ser superior a la poblaci\'on.

\section{Relevancia del parametro $\beta^{*}-\gamma$}

A continuaci\'{o}n discutiremos las soluciones del sistema SIR en los casos de
que el n\'{u}mero de infectados es mucho m\'{a}s peque\~{n}o que la
poblaci\'{o}n $I\ll N$. Esta aproximaci\'{o}n es relevante con vistas a
discutir los casos de los pa\'{\i}ses que han superado la epidemia, en estos
pa\'{\i}ses esta relaci\'{o}n se cumple. En este caso el n\'{u}mero de
susceptibles se puede aproximar por la misma poblaci\'{o}n del pa\'{\i}s
$S(t)\approx N$.

Por tanto el sistema de ecuaciones (\ref{sir1},\ref{sir2},\ref{sir3}) se
simplifica a la forma
\begin{align}
\frac{\partial S}{\partial t} &  =-\beta N\text{ }I,\\
\frac{\partial I}{\partial t} &  =\beta\text{ }NI-\gamma\text{ }I,\\
\frac{\partial R}{\partial t} &  =\gamma\text{ }I.
\end{align}
Dado que las ecuaciones son lineales, las soluciones resultan exponenciales
dadas por:
\begin{align}
S_{e}(t) &  =N-\beta^{\ast}\text{ }i\text{ }Exp((\beta^{\ast}-\gamma)t),\\
I_{e}(t) &  =\text{ }i\text{ }Exp((\beta^{\ast}-\gamma)t)\\
R_{e}(t) &  =\frac{\gamma}{\beta^{\ast}-\gamma}\text{ }i\text{ }%
Exp((\beta^{\ast}-\gamma)t).\\
\beta^{\ast} &  =\beta N,
\end{align}
donde $i$ constituye el \'{u}nico par\'{a}metro libre de la soluci\'{o}n. Para
pa\'{\i}ses con millones de habitantes, como antes se mencion\'{o}, la
soluci\'{o}n exponencial es una muy buena aproximaci\'{o}n para tiempos en que
$I(t)\ll N.$

 En este punto es de inter\'es subrayar que si en esas ecuaciones se considera $\beta^*$ como una variable
 dependiente del tiempo, el signo de $\beta^*(t)-\gamma$ determina el signo de la derivada temporal de $I(t)$.
 Es decir,  que la magnitud $\beta^*(t)-\gamma$ brinda un factor tan relevante  como si la epidemia crece o decrece
 en el instante considerado.

En el caso que no hay contagio $\beta=0$, por lo cual el valor de $\gamma$
corresponde predecir un decaimiento exponencial de los infectados (caso de
China). Sin embargo el tiempo t\'{\i}pico en que se cura cada enfermo es de 15
a 20 d\'{\i}as. Por lo tanto, adoptaremos en esta secci\'{o}n el estimado de
$\gamma=\frac{1}{15}=0.066.$
Tal como comentamos en la Introducci\'{o}n, las soluciones exponenciales
tienen un tiempo caracter\'{\i}stico de crecimiento o decrecimiento
determinado por el n\'{u}mero $\beta^{\ast}-\gamma$. Si esta cantidad resulta
positiva la soluci\'{o}n del problema crece indefectiblemente,
independientemente de los valores especficos de $I$ y $R$ en el momento dado.
Esta es una propiedad relevante en la din\'{a}mica de este problema. Para que
$I$ decrezca en la inmportante regi\'{o}n de los bajos valores es
imprescindible que el valor en el tiempo de $\beta^{\ast}-\gamma$ resulte
negativo. De acuerdo a esto, en todos los pa\'{\i}ses en que se ha logrado la
recuperaci\'{o}n de la epidemia, se ha podido hacer que esa cantidad sea
negativa, dando un m\'{a}ximo de infectados activos con $I\ll N$.

Sin embargo,
es tambi\'{e}n conocido que aunque el pa\'{\i}s imponga aislamiento total a
partir de cierta fecha, las curvas de infecci\'{o}n en ning\'{u}n caso
comienzan a bajar instantaneamente. Es decir, aunque es l\'{o}gico suponer las
medidas deben fijar instantaneamente la condici\'{o}n $\beta^{\ast}=0$
(ausencia de transmisibilidad) los datos de infecci\'{o}n niegan esta
propiedad y las pendientes continuan siendo positivas d\'{\i}as despu\'{e}s de
la fecha de confinamiento. Analicemos las causas de este comportamiento en la
siguiente secci\'{o}n.

\section{Modelo cualitativo del aislamiento}

Describamos ahora un modelo cualitativo que brinda una raz\'{o}n por la que el
valor de $\beta^{\ast}=0$ no se hace cero inmediatamente despu\'es de tomar
las medidas de aislamiento. Consideremos la circunstancia de que el tiempo en
que los enfermos sufren el padecimiento se estima en un per\'iodo $\tau=15-20$
d\'{\i}as. El enclaustramiento de las familias, implica que en gran cantidad
de ellas existen enfermos asintom\'{a}ticos o infectados leves que pueden
transmitir la enfermedad a sus familiares. Por tanto se puede estimar que el
n\'umero positivo de susceptibles que pasan a ser infectados por unidad de
tiempo no se puede anular al instante de implantar el aislamiento. Por tanto
$\beta^{\ast}(t)$ no debe reducirse lo suficiente para cambiar el signo de
$\beta^{\ast}-\gamma$ ($\tilde R_{0}<1$). Sin embargo, lo que debe ser
v\'alido es que tras un per\'iodo cercano al tiempo de vida de la enfermedad,
ya no deben existir casos de enfermos leves en las familias. Por lo cual
despu\'es de ese intervalo de tiempo $\beta^{\ast}\rightarrow0$. El modelo
propuesto se basa centralmente en esta afimaci\'on.

Es decir, que un d\'{\i}a despu\'{e}s de la implantaci\'{o}n del
confinamiento, y durante digamos aproximadamente veinte d\'{\i}as $\beta
^{\ast}$ debe decrecer hasta anularse. La dependencia temporal en ese
intervalo no es conocida. Solo es de esperar que la funci\'{o}n tenga una
disminuci\'{o}n brusca el preciso d\'{\i}a en que comienza el aislamiento,
dado que las condiciones de contacto de los infectados con su entorno
cambiaron dr\'{a}sticamente. Una vez que la transmisibilidad se anula, la
curva de deca\'{\i}da del n\'{u}mero de infectados $I(t)$ debe tener un
car\'{a}cter exponencial con tiempo de decaimiento $\gamma.$ Como antes
mencionamos la curva de infecci\'{o}n de China reportada en el sitio web
www.worldometer.info \cite{11}, permiten estimar el valor de $1$/$\gamma$
entre $15$ y $20$ d\'{\i}as.

Haremos un comentario final acerca de las suposiciones del modelo. Considerar
$\beta^{*}=0$ luego de un tiempo de duraci\'on de la afecci\'on despu\'es de
instaurado el aislamiento, es v\'alido si se supone que este aislamiento es
unipersonal. Es decir si cada persona se encuentra aislada. Pasado este
intervalo, el n\'umero de susceptibles no puede variar con el tiempo pues no
es posible infectarlos. Siendo Alemania uno de los pa\'{\i}ses m\'as
desarrollados, esa condici\'on pudiera satisfacerse aproximadamente. Como se ver\'a en la
siguiente subseccci\'on, el modelo funciona bastante bien para ese pa\'is. Sin
embargo, en otras situaciones puede suceder que al final de un per\'iodo de
duraci\'on despu\'es del confinamiento, la tasa de contagio comience a
decrecer a partir de un valor disminu\'ido. Esto es de esperar en pa\'{\i}ses
donde el aislamiento se aleje bastante del individual, donde existe
aglomeraci\'on de personas en la familia media. En esos casos, supondremos que
pasado el citado intervalo de tiempo los valores de $\beta^{*}$ disminuyen
exponencialmente con una constante de decaimiento cuyo valor controlar\'a el
m\'aximo de $I_{o}(t)$. A continuaci\'on el modelo cualitativo descrito de la
evoluci\'{o}n de $\beta^{\ast}$ bajo condiciones de aislamiento se aplica a
describir las curvas de infectados de Alemania y Cuba.

\subsection{Aplicaci\'on a Alemania y a Cuba}

Consideremos en esta subsecci\'{o}n una descripci\'{o}n de las curvas de
infecci\'{o}n de Alemania y Cuba en base al modelo simple descrito. Para ello,
primeramente fijamos los par\'{a}metros $\beta^{*}$ y $\gamma$ en el comienzo
de crecimiento exponencial de la infecci\'{o}n en ambos pa\'{\i}ses. El valor
de la relaci\'on $k$ entre el n\'umero de infectados observados y el total se
tom\'o como $k=0.2.$

\subsubsection{Alemania}

Discutamos primeramente el caso de Alemania. Para este pa\'{\i}s se tiene una
importante informaci\'{o}n: sus medidas de aislamiento comenzaron a imponerse
fuertemente cerca del 20 de marzo del 2020. Los datos de curvas de infectados
al inicio de la epidemia permitieron estimar $\beta^{\ast}$ y \ $\gamma$ cuyos
valor resultaron:
\[
\beta^{\ast}=0.31\,\,,\text{ \ }\gamma=1/20.
\]
El valor de $\gamma$ se estim\'{o} de los datos de la ca\'{\i}da exponencial
de la cantidad de infectados  en China \cite{11}.

\begin{figure}[h]
\begin{center}
\includegraphics[width=.6\textwidth]{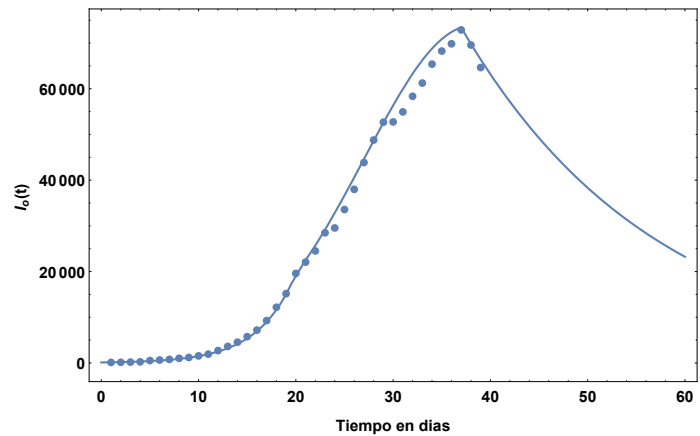}
\end{center}
\caption{{\protect\footnotesize El gr\'{a}fico muestra en la curva de puntos
los datos para el n\'{u}mero de infectados como funci\'{o}n de los d\'{\i}as
para Alemania \cite{11}. La l\'{\i}nea continua muestra la soluci\'{o}n del
modelo SIR. Los valores de $\beta^{\ast}$ y $\gamma$ son constantes en los
primeros 20 d\'{\i}as, en que las medidas de aislamiento no hab\'{\i}an sido
impuestas. N\'{o}tese que la curva continua
{\protect\footnotesize (exponencial en la aproximaci\'{o}n lineal) }sigue muy
exactamente a la observada en ese intervalo. En el rango 20-40 d\'{\i}as la
variaci\'{o}n de la funci\'{o}n }${\protect\footnotesize \beta}^{\ast}%
${\protect\footnotesize  se muestra en la figura \ref{beta}. Note que la
ca\'{\i}da exponencial del n\'{u}mero de infectados observados comienza
aproximadamente un per\'{\i}odo de duraci\'{o}n de la enfermedad despu\'{e}s
adoptado el aislamiento. }}%
\label{Alemania}%
\end{figure}

Estos par\'{a}metros determinaron una dependencia exponencial para cortos
tiempos que describe muy bien la curva de datos reportados para Alemania en la
referencia \cite{11}, hasta los 20 d\'{\i}as del mes de marzo. En el entorno
de esta fecha fueron impuestas las medidas de aislamiento. La curva de
$I_{o}(t)$ en esa regi\'{o}n de tiempos se muestra en la figura \ref{Alemania}.

\begin{figure}[h]
\begin{center}
\includegraphics[width=.6\textwidth]{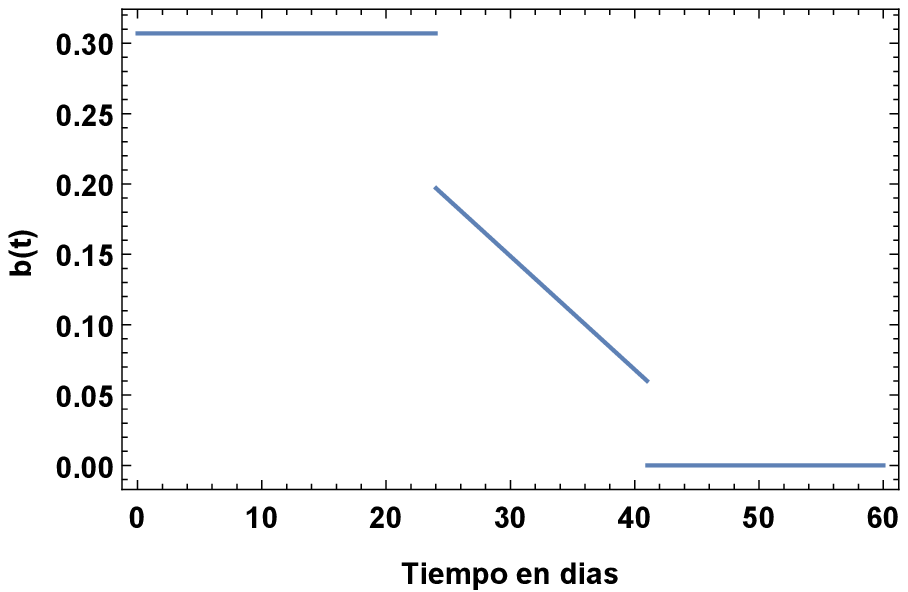}
\end{center}
\caption{ {\protect\footnotesize Se muestra la dependencia temporal de
$\beta^{\ast}$ que reproduce la curva de infectados de Alemania. Una primera
disminuci\'{o}n brusca  al imponer el aislamiento  es de esperar debido a que
el confinamiento reduce las posbilidades  para el contagio con cada infectado.
El salto al valor nulo tambi\'{e}n puede esperarse debido a que transcurrido
el tiempo medio de infecci\'{o}n, es natural que no haya casi disminuci\'{o}n
del n\'{u}mero de susceptibles por unidad de tiempo, o sea $\beta^{\ast}$ sea
cercano a cero. La variaci\'{o}n en la zona interior se disminuy\'{o}
linealmente y la pendiente se determina con vistas a un ajuste de los datos en
la zona de sus m\'{a}ximos valores. }}%
\label{beta}%
\end{figure}

A continuaci\'{o}n se utiliza que la curva de infecci\'{o}n muestra un cambio
en pendiente ese mismo d\'{\i}a 20 de marzo. Esto permiti\'{o} estimar el
nuevo valor de $\beta^{\ast}$ al que disminuye esta magnitud bruscamente, al
imponer el aislamiento. Posteriormente, dado que no se tiene informaci\'{o}n a
priori acerca del comportamiento de $\beta^{\ast}$ en los 20 d\'{\i}as (tiempo
medio que dura la enfermedad) posteriores al 20 de marzo, \ se asumi\'{o} que
$\beta^{\ast}$ disminuye linealmente durante el tiempo de permanencia de la
enfermedad hasta que bruscamente salta al valor nulo tras esos 20 d\'{\i}as.
Esto implementa el modelo de la secci\'{o}n anterior. La figura \ref{beta}
muestra esta variaci\'{o}n. Ajustando la  dependencia lineal  de $\beta^{\ast
}$ para tiempos posteriores al instante de aislamiento, se logra entonces
describir satisfactoriamente la curva de infecci\'{o}n de Alemania, como lo
muestra la figura \ref{Alemania}.

Puede decirse que el caso de Alemania brinda un ejemplo de que la correcta
implantaci\'{o}n de las medidas de aislamiento puede determinar un decaimiento
exponencial de la pandemia, transcurrido aproximadamente un per\'{\i}odo medio
de duraci\'{o}n de la enfermedad. Pensamos que este efecto pudiera ser
relevante para explicar los casos de otros pa\'{\i}ses en esta etapa como
Corea del Sur, Austria, Suiza, etc.

\subsubsection{Cuba}

Posteriormente el mismo modelo se aplic\'{o} a la curva de infecci\'{o}n de
Cuba, asumiendo la fecha del 24 de marzo 2020 para la imposici\'{o}n del
aislamiento. Tomando $\gamma=0.05$, para el par\'{a}metro $\beta^{\ast}$
derivado de los datos diarios de infectados brindados en la referencia
\cite{11}, se obtuvieron los valores
\[
\beta^{\ast}=0.307088,\,\,\gamma=1/20.
\]

\begin{figure}[h]
\begin{center}
\includegraphics[width=.6\textwidth]{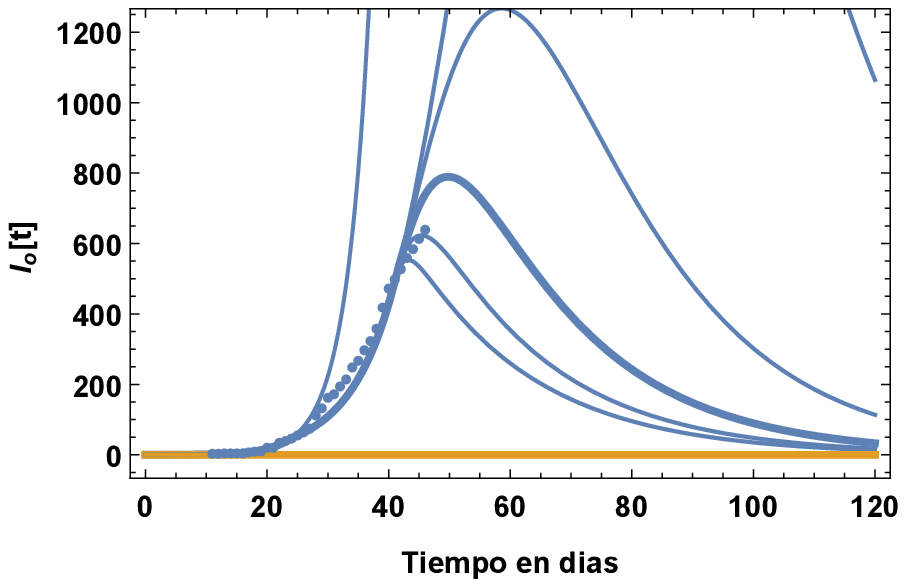}
\end{center}
\caption{ {\protect\footnotesize La figura muestra la soluci\'{o}n del sistema
de ecuaciones SIR para Cuba resuelto utilizando un coeficiente $\beta^{\ast}$ constante
antes del d\'{\i}a 24 de marzo en que se implementa el aislamiento de la
poblaci\'{o}n. Este valor ajusta los datos de n\'{u}mero de infectados
activos como funci\'{o}n del tiempo en ese intervalo inicial. M\'{a}s all\'{a}
del citado d\'{\i}a, se considera que $\beta^{\ast}$ decrece, mostrando una
disminuci\'{o}n brusca a un valor constante durante un tiempo medio de
duraci\'{o}n de la afecci\'{o}n $\tau$. Al final de
ese intervalo, se supone que $\beta^{\ast}$ decae exponencialmente de acuerdo
a  $exp(-\alpha(t-24-\tau ))$. Esto se  adopta para
describir el esperado decaimiento a cero de $\beta^{\ast}$%
  tras el aislamiento. \ Las cinco cuervas muestran las
soluciones para el n\'umero de infectados para los valores $\alpha
=0.6,0.3,0.15,0.075,0.0375.$ El orden de estas constantes  indica las curvas
con valores crecientes de sus m\'aximos valores.}}%
\label{Cuba}%
\end{figure}

Los resultados para $I_{o}(t)$ entre el 1 de marzo (fecha en que se comienza
medir el tiempo $t$ en d\'ias) y el 24 Marzo, se muestran a la izquierda de la
figura \ref{Cuba}. En ella los puntos presentan  los datos reportados por el
Ministerio de Salud P\'{u}blica y las curvas continuas representan las
soluciones del sistema de ecuaciones SIR. Se puede observar que la curva
continua describe aceptablemente los datos del n\'{u}mero de infectados antes
de establecerse el aislamiento el 24 de marzo ($t=24$). Cabe subrayar que la
curva exponencial a la extrema izquierda en el gr\'{a}fico, es la soluci\'{o}n
de las ecuaciones SIR en ausencia de las condiciones de aislamiento. Ella indica
que en el caso de no haberse aplicado el confinamiento, el crecimiento del
n\'{u}mero de infectados hubiera sido muy grande, dos semanas despu\'{e}s del
24 de Marzo, mostrando claramente el papel del aislamiento. \  Posteriormente al 24 de marzo, se supuso, tal como en el
modelo, que $\beta^{\ast}$ disminuye bruscamente precisamente ese d\'{\i}a, a
un valor, menor. En este caso supusimos que dicho valor permanece constante
tras 17 de d\'{\i}as de duraci\'{o}n t\'{\i}pica de la afecci\'{o}n.
Transcurrido ese tiempo, consideramos que $\beta^{\ast}$ tiende cero, pero no
bruscamente. En lugar de esto, asumiremos que tiende a cero con una
dependencia exponencial del tipo%
\[
\beta^{\ast}(t)=Exp[-\alpha\text{ }(t-(24+\tau))],
\]
donde en este caso,  tomamos $\tau=17$ como el tiempo de medio de la enfermedad.
Las  curvas que se muestran en la figura \ref{Cuba} para tiempos mayores
que $24+\tau$ corresponden a los valores de $\alpha
=0.6,0.3,0.15,0.075,0.0375.$ Estas consideran seis formas de decaimiento exponencial
de $\beta^{\ast}$ con vistas a intentar describir la  esperada disminuci\'on de $\beta^{\ast}$ a
suceder despu\'es de establecido el aislamiento y esperar un tiempo de duracion
de la enfermedad. Como se  coment\'o en la discusi\'on del modelo, en pa\'ises en
que el  confinamiento se logra con mas dificultad,  solo cabe esperar que
$\beta\ ^{\ast}$ decaiga despues de esperar un tiempo $\tau$ a partir del d\'ia
de aislamiento.  Es inter\'es entonces comparar las observaciones que se
ofrecen por el Ministerio de Salud P\'{u}blica con las curvas de infectados
para varios  valores del parametro $\alpha$.
Obs\'{e}rvese que para $\alpha=0.6$ la exponencial se reduce bastante en el
curso de dos d\'{\i}as, por lo que para ese valor se puede considerar que se
est\'{a} imponiendo $\beta^{\ast}$ $=0$ una vez pasado un per\'{\i}odo medio
de duraci\'{o}n de la enfermedad.  Esto puede considerarse como una buena
aproximaci\'on de una respuesta similar a la de Alemania.  Para valores mayores
de $\alpha$ las curvas de la figura predicen m\'aximos de mayor valor que van
apareciendo a  tiempos tambi\'en superiores.  Los tiempos a que ocurren est\'an
descritos por los ceros de la funci\'on
\begin{equation}
\frac{I^{\prime}(t)}{I(t)}=\beta^{\ast}(t)-\gamma,
\end{equation}
que se plotea en la figura \ref{beta1}.  Es claro de los datos observados,
que los dos primeros ma\'ximos, asociados a $\alpha=0.6$ $\ $y $0.3$\  no pueden
ocurrir,  pues los datos de n\'umero de infectados han sobrepasado esos valores
ya.  Los datos de 16 de abril se acercan a los de la curva de  $\alpha=0.15$
pero no  muestran a\'un una derivada cercana  a cero, por lo que su m\'aximo
(de cerca de 800 casos activos )  no
debe realizarse tampoco.  Para  $\alpha=0.075,$ sin  embargo el m\'aximo
 se aparecer\'ia para cerca del 1 de mayo,  mostrando 1200 casos . Por otro lado la curva de
$\alpha=0.0375$  mostrar\'ia su m\'aximo cerca del 12 de mayo con alrededor de
2000 casos de infectados activos

Cabe estimar  que dados lo valores de las observaciones mostradas, existen
buenas posibilidades  para que los m\'aximos a ocurrir realmente  est\'en  por debajo de
1,000 o  2, 000 casos

La variaci\'{o}n temporal descrita para $\beta^{\ast}$ se muestra en la figura
\ref{beta1} para el caso de $\alpha=0.15$. La curva de $I_{o}(t)$ asociada es
la l\'{\i}nea continua m\'{a}s gruesa en la figura \ref{Cuba}.
\begin{figure}[h]
\begin{center}
\includegraphics[width=.6\textwidth]{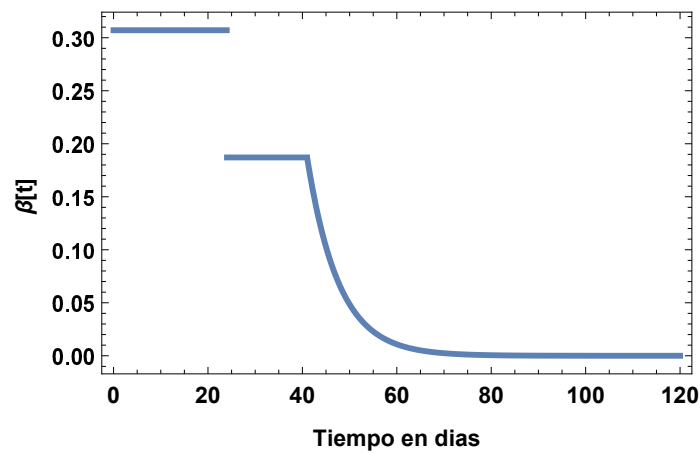}
\end{center}
\caption{ {\protect\footnotesize El gr\'{a}fico muestra la variaci\'{o}n
temporal de $\beta^{\ast}$ que permite describir la curva de n\'{u}mero de
infectados observados m\'{a}s gruesa que se muestra en la figura \ref{Cuba}.
Esta corresponde a $\alpha=0.15$. }}%
\label{beta1}%
\end{figure}\

Como ya mencionamos, la figura \ref{beta1} muestra las curvas de $I^{\prime
}(t)/I(t)=\beta^{\ast}(t)-\gamma$ para el modelo y para los datos, estimada
para estos por la raz\'{o}n entre el n\'{u}mero de enfermos reportados
diariamente y la cantidad total de enfermos activos. A pesar de las
fluctuaciones, debidas a que la cantidad de casos es a\'{u}n reducida (respecto
a la poblaci\'{o}n), se indentifica cierta coherencia con la dependencia
temporal de la $\beta^{\ast}(t)-\gamma$ asumida en el marco del modelo considerado.
La dependencia m\'as all\'{a} del d\'{\i}a 16 de abril, en que se t\'ermina de
redactar este trabajo, debe decidir cuan v\'{a}lida es la suposici\'{o}n
sugerida por el modelo de Alemania: considerar que $\beta^{\ast}(t)$ debe tender a
cero despu\'{e}s de un tiempo de duraci\'{o}n medio de la enfermedad.

Una sugerencia importante que puede extraerse de las figuras \ref{beta1} \ y
\ref{Cuba} se refiere a asumir que la curva azul de los datos para $I^{\prime
}(t)/I(t)$ en \ref{beta1} se debe esperar que se apegue a alguna de las
exponenciales ploteadas en ella. Si ello es as\'i, se sigue  que las cotas de
un m\'aximo no superior a 1,000 o 2,000 casos, a suceder antes del 12 de mayo,
 pueden resultar  v\'alidas como indica la
figura  \ref{Cuba}.

\begin{figure}[h]
\begin{center}
\includegraphics[width=.6\textwidth]{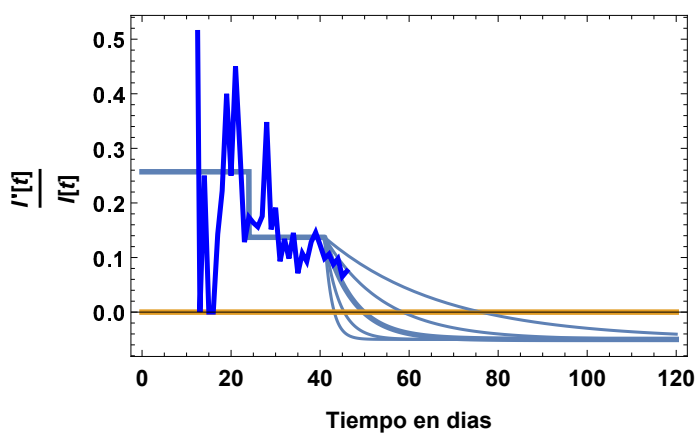}
\end{center}
\caption{ {\protect\footnotesize Se ilustran las curvas de $I^{\prime
}(t)/I(t)=\beta^{\ast}(t)-\gamma$ para la soluci\'{o}n y para los datos,
estimada para estos por la raz\'{o}n entre el n\'{u}mero de enfermos
reportados diariamente y la cantidad total de ellos que permanecen enfermos
dicho d\'{\i}a (enfermos activos). El gr\'{a}fico muestra cierta semejanza
entre la $\beta^{\ast}(t)$ asumida en el marco del modelo con la observada.
Las 5 curvas exponenciales que aparecen corresponden a los varios valores de
}$\alpha=0.6,0.3,0.15,0.075,0.0375.${\protect\footnotesize La curva gruesa
corresponde a $\alpha=0.15$ . Esta  figura y la figura \ref{beta1} sugieren
que el m\'aximo del n\'umero de casos podr\'ia estar  acotado entre 1,000 y 2,000, si
\ la curva azul de }$I^{\prime}(t)/I(t)${\protect\footnotesize  observado
corta el eje horizontal antes del 12 de mayo. }}%
\label{beta1}%
\end{figure}\

\newpage

\subsection{Ajustes a modelos con tasa de contagio variable}

 \begin{table}[t]
\begin{center}%
\begin{tabular}
[c]{|c|c|c|c|}\hline
Per\'{\i}odos & $\gamma(\tilde R_{0}-1)$ & $\beta^{*}$ & $\tilde R_{0}%
$\\\hline
I: 16.03.20 \text{ al } 21.03.20 & 0.38313 & $0.43313$ & 8.66261\\\hline
II: 22.03.20 \text{ al } 28.03.20 & 0.207859 & $0.257859$ & 5.15717\\\hline
III: 29.03.20 \text{ al } 9.04.20 & 0.105325 & $0.155325$ & 3.10649\\\hline
IV: 10.04.20 \text{ al } 16.04.20(\text{ o en adelante}) & 0.0460772 &
$0.0960772$ & 1.92154\\\hline
V: 30.04.20-\text{ en adelante} & -0.0115691 & $0.0384309$ & 0.768618\\\hline
\end{tabular}
\end{center}
\caption{{\protect\footnotesize Valores de tasa de contagio variables de Cuba
obtenidas ajustando localmente al comportamiento exponencial del SIR para
tiempos peque\~{n}os. En el primer escenario se extrapola con la tasa de
contagio del periodo IV. En el segundo escenario se fija una tasa de contagia
reducida a partir del periodo V. Se considera una tasa de recuperaci\'{o}n
$\gamma=1/20$, estimada de la cola de la curva de China. }}%
\label{pedazos}%
\end{table}

En esta subsecci\'{o}n realizaremos un estudio experimental del cambio de la
tasa de contagio $\beta$ para varios pa\'{\i}ses. Estos resultados los
comparamos con el modelo presentado anteriormente, llegando a la
conclusi\'{o}n de que efectivamente esta tasa de contagio variable se ajusta a
los datos, y se reduce con el proceso de cuarentena. Por lo cual el estudio de
la din\'{a}mica de esta tasa de contagio es relevante.
\begin{figure}[h]
\begin{center}
\includegraphics[width=.45\textwidth]{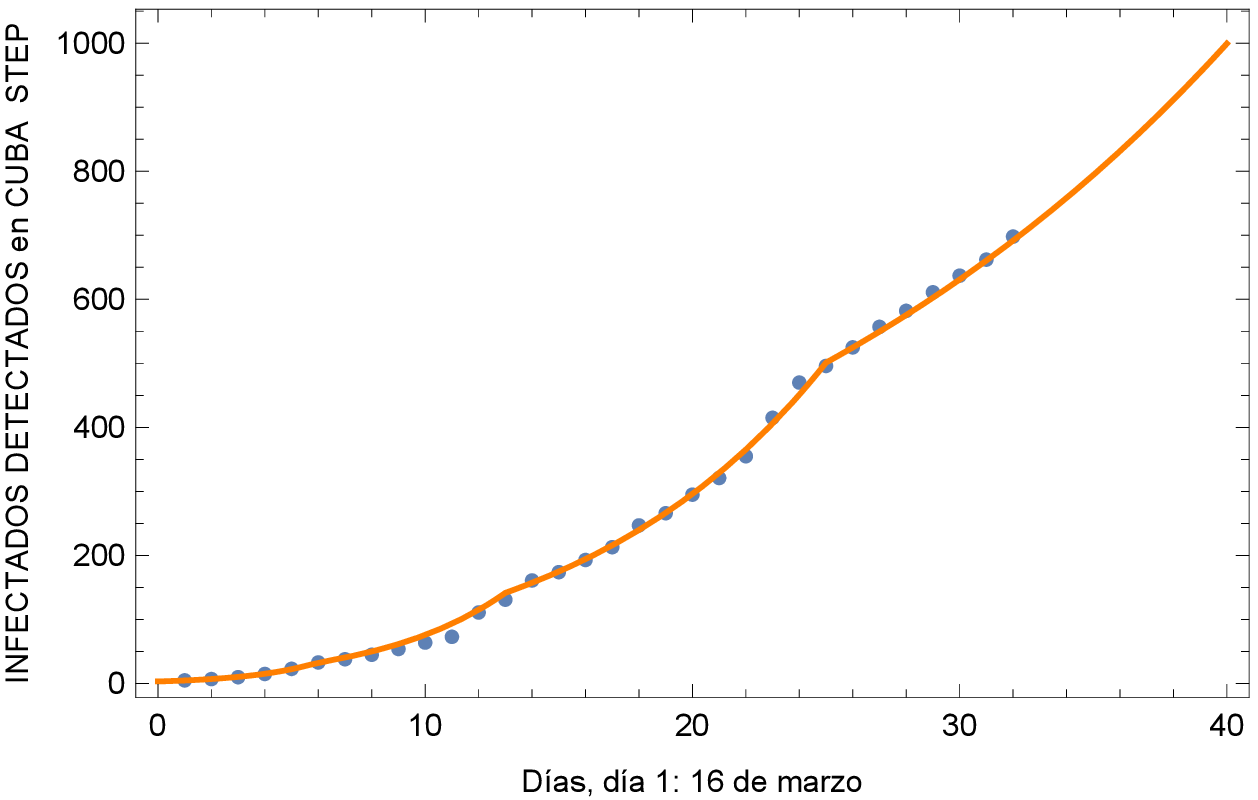}
\includegraphics[width=.45\textwidth]{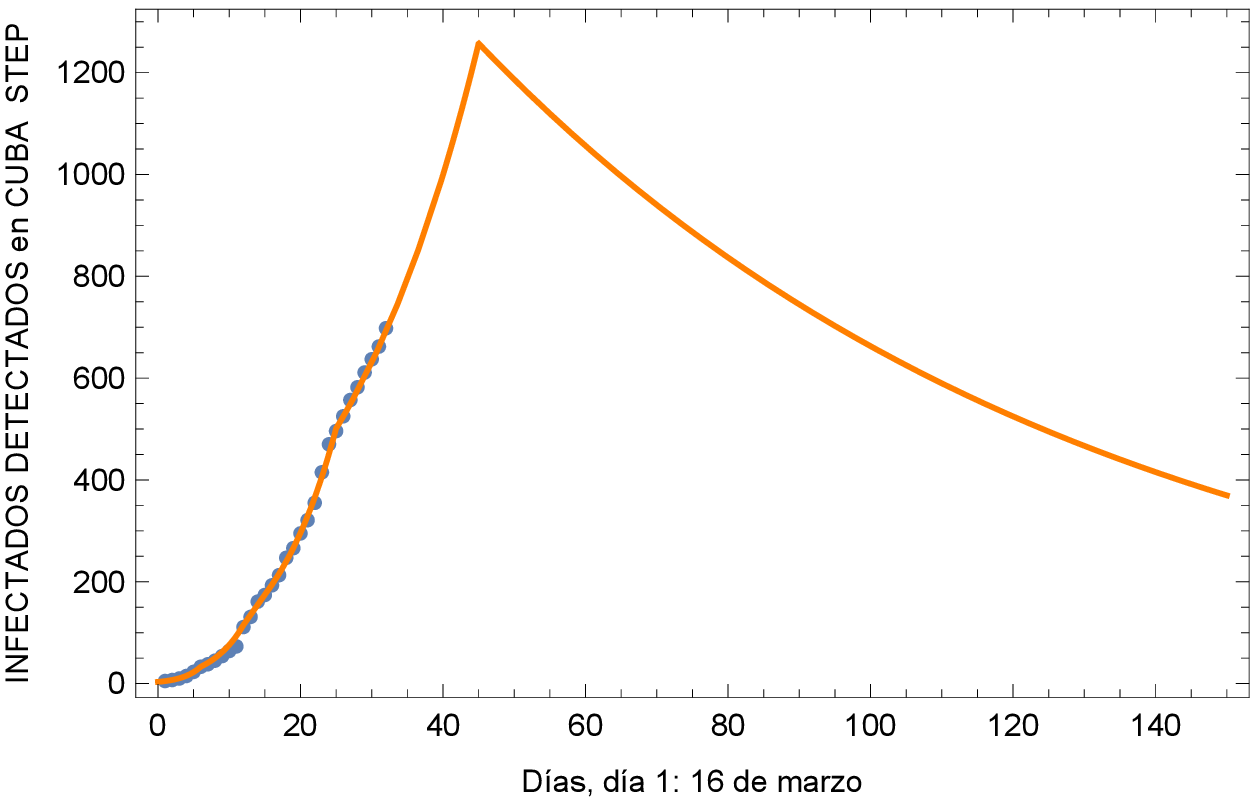}
\end{center}
\caption{{\protect\footnotesize El panel de la izquierda representa el
n\'{u}mero de infectados total dado en puntos, la curva naranja es la
predicci\'{o}n del modelo SIR, con una taza de contagio variable. El panel
derecho  representa el caso optimista en que el confinamiento logre alcanzar
una tasa de contagio efectiva con  $\protect\widetilde{R}_{0}<1$ a finales de
abril. Los datos considerados corresponde  al 18.04.20. }}%
\label{ajusteE}%
\end{figure}

Quisi\'{e}ramos destacar adem\'{a}s que estudiando el caso de Alemania y el
caso de Corea del Sur comprobaremos que el m\'{a}ximo alcanzado (y deseado) se
corresponde con un m\'{a}ximo debido a $\beta^{\ast}-\gamma<0$, o lo que es lo
mismo $\tilde{R}_{0}<1$. El cual no es el m\'{a}ximo estándar del SIR, si no
una logrado en la regi\'{o}n donde $S\approx N$.
\begin{figure}[h]
\begin{center}
\includegraphics[width=.45\textwidth]{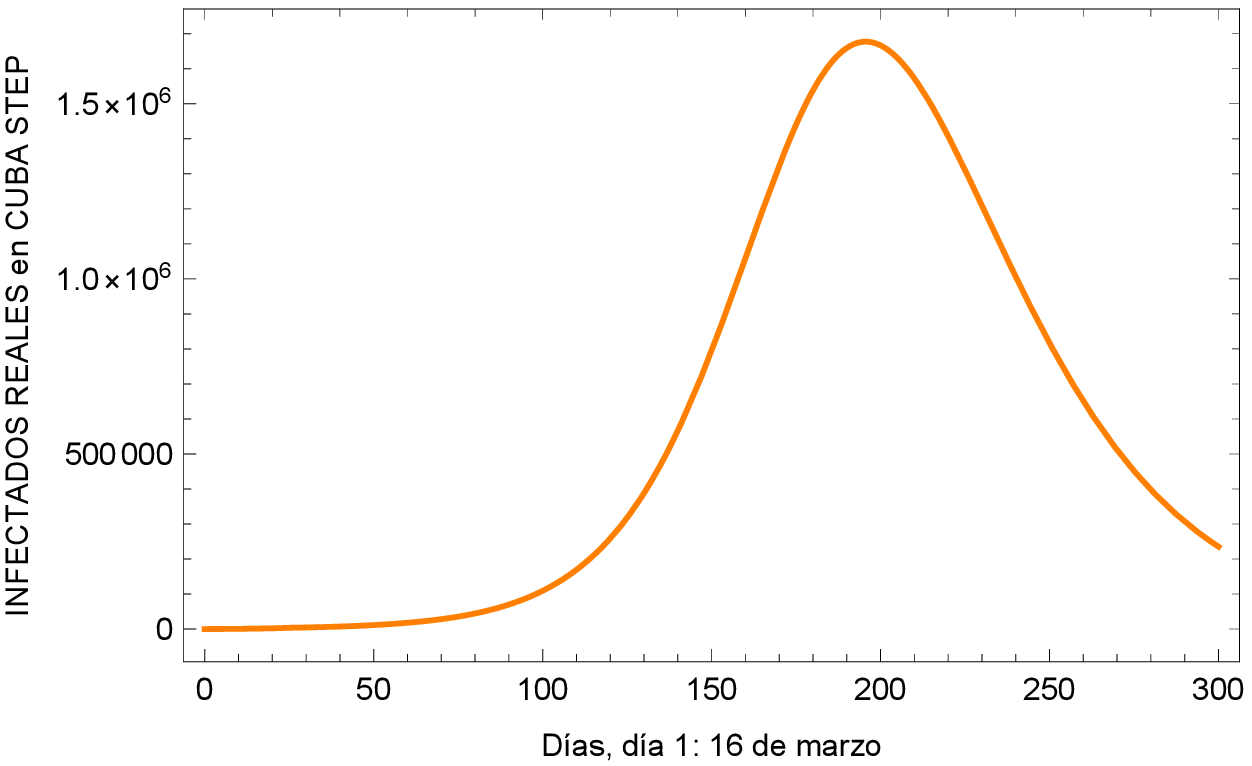}
\includegraphics[width=.45\textwidth]{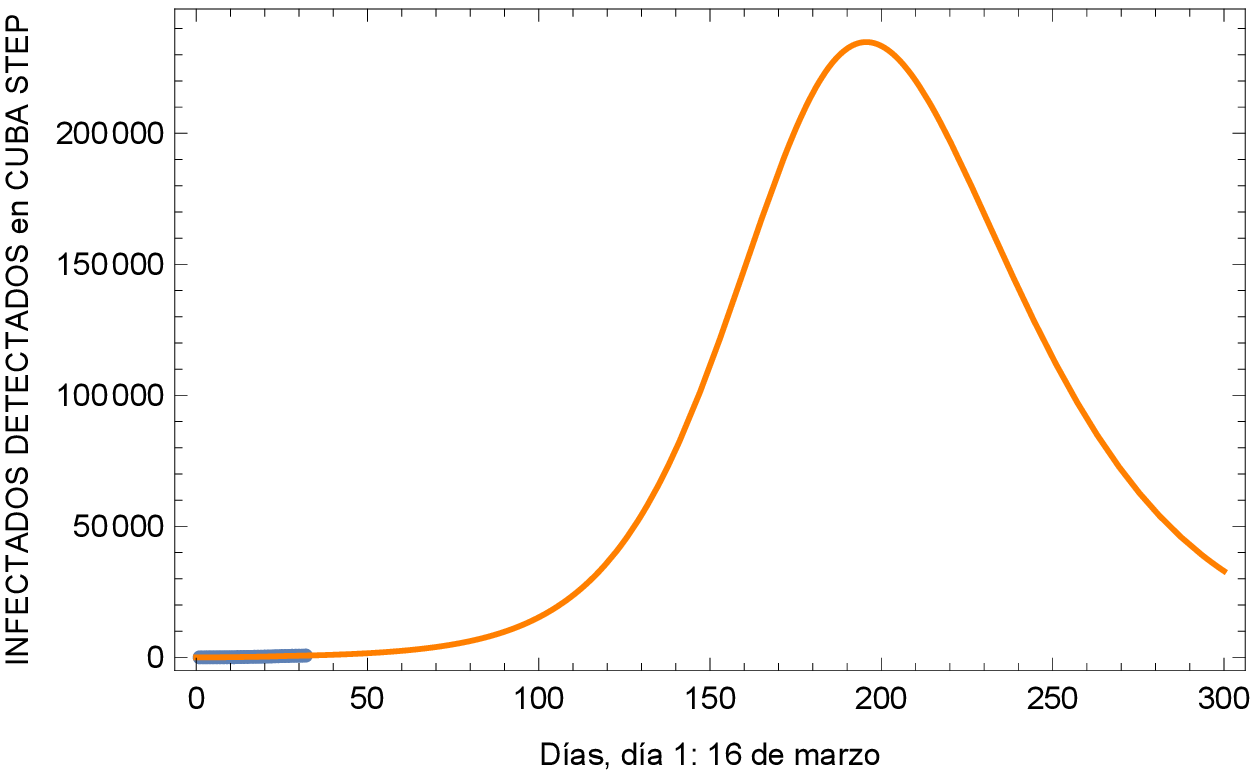}
\includegraphics[width=.45\textwidth]{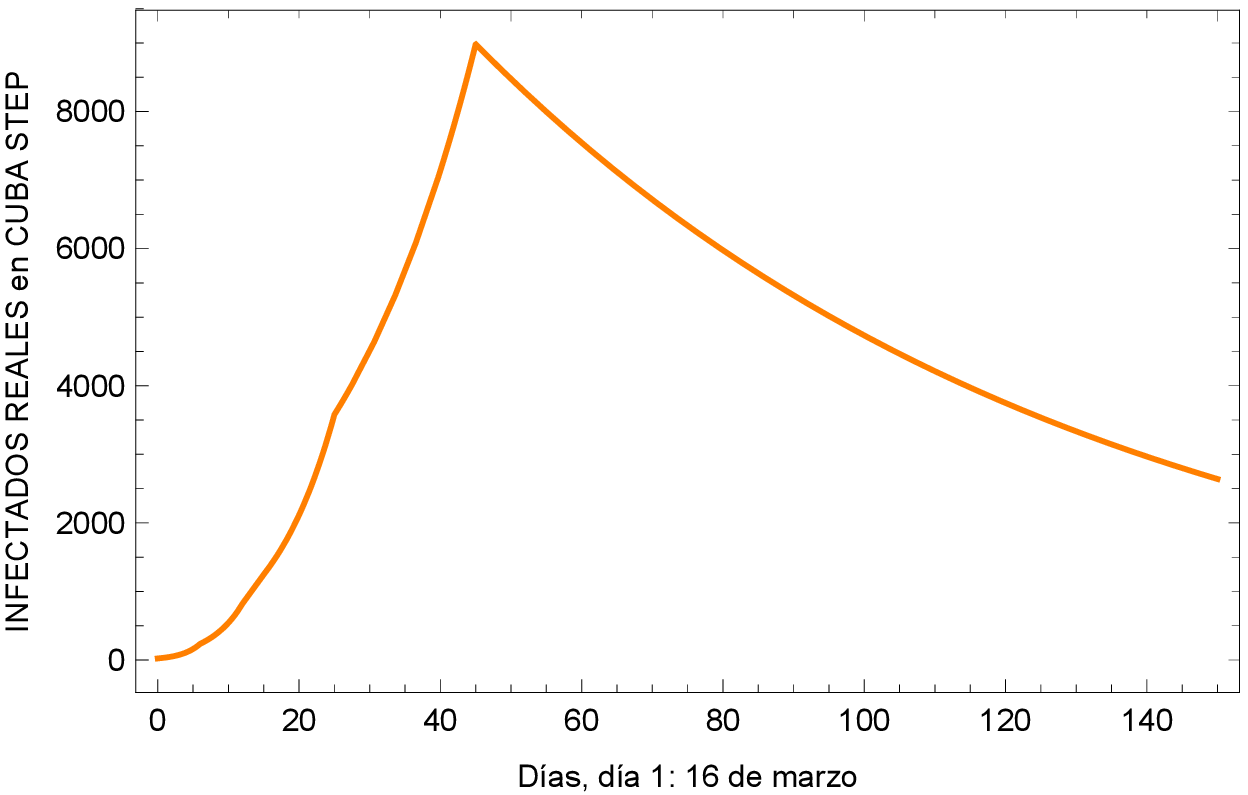}
\includegraphics[width=.45\textwidth]{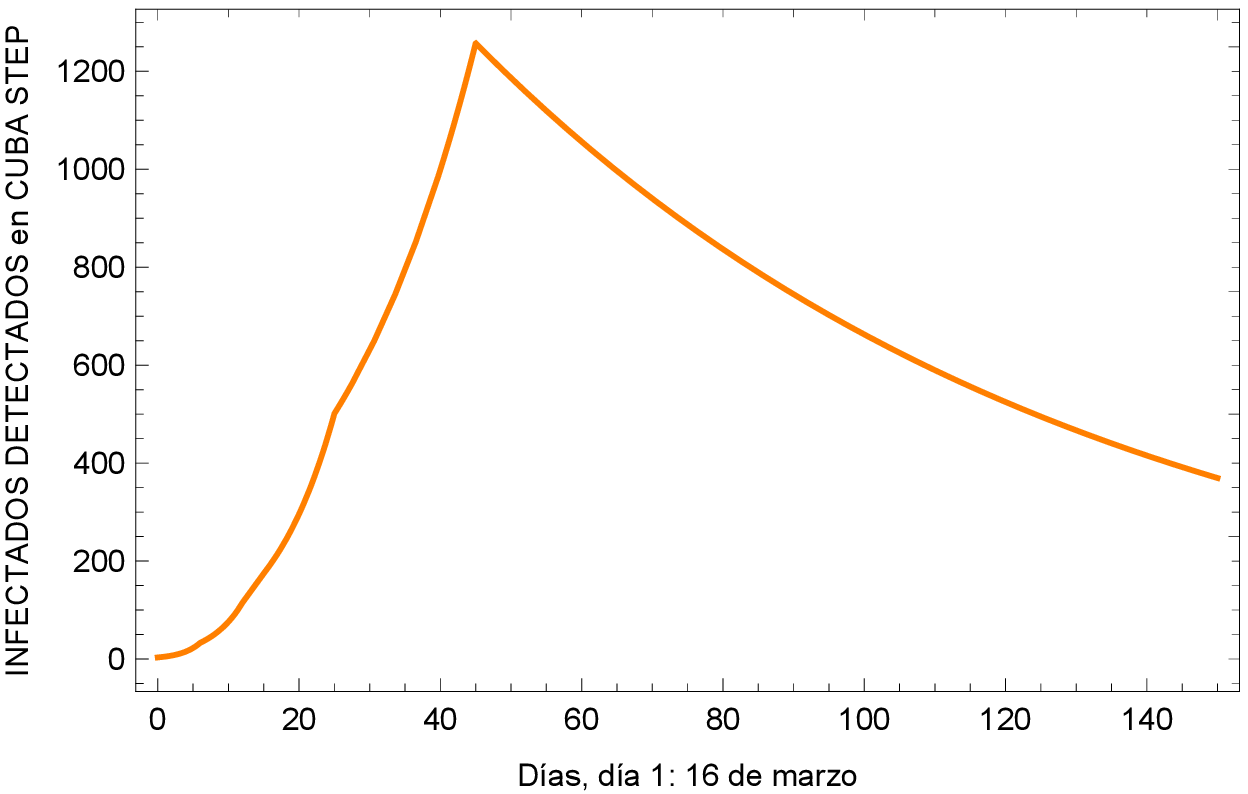}
\end{center}
\caption{{\protect\footnotesize N\'{u}mero de infectados predichos por el
modelo SIR en funci\'{o}n del tiempo. Se considera una taza de detecci\'{o}n
de $k=0.14$ \cite{5,12}. En la primera fila, se dan los graficos de infectados
totales y observados, respectivamente, considerando una tasa de contagio
variable, ajustada a datos del pa\'{\i}s. La extrapolaci\'{o}n a grandes
tiempos se hace empleando la tasa de contagio al 16.04.20. La segunda fila
muestra la evoluci\'{o}n \ de infectados totales y observados \ para  el caso
de imponer un confinamiento m\'{a}s estricto a finales de abril (tabla).}}%
\label{ajusteE2}%
\end{figure}
Empleando los datos que se tienen hasta el 16.04.20 construimos un modelo SIR
para Cuba dividido en cuatro etapas, considerando valores de $\beta^{\ast}$
dependientes del tiempo. Los valores se muestran en la tabla \ref{pedazos}.
Estos valores de $\beta^{\ast}$ locales se emplean para construir dos
escenarios, el primero con tres valores de $\beta^{\ast}$ y extrapolando la
evoluci\'{o}n para tiempos mayores con la \'{u}ltima tasa de contagio calculada.
El segundo, suponiendo que a finales de abril el $\beta^{\ast}$ se
reduce dando $\tilde{R}_{0}<1$.
\begin{figure}[h]
\begin{center}
\includegraphics[width=.45\textwidth]{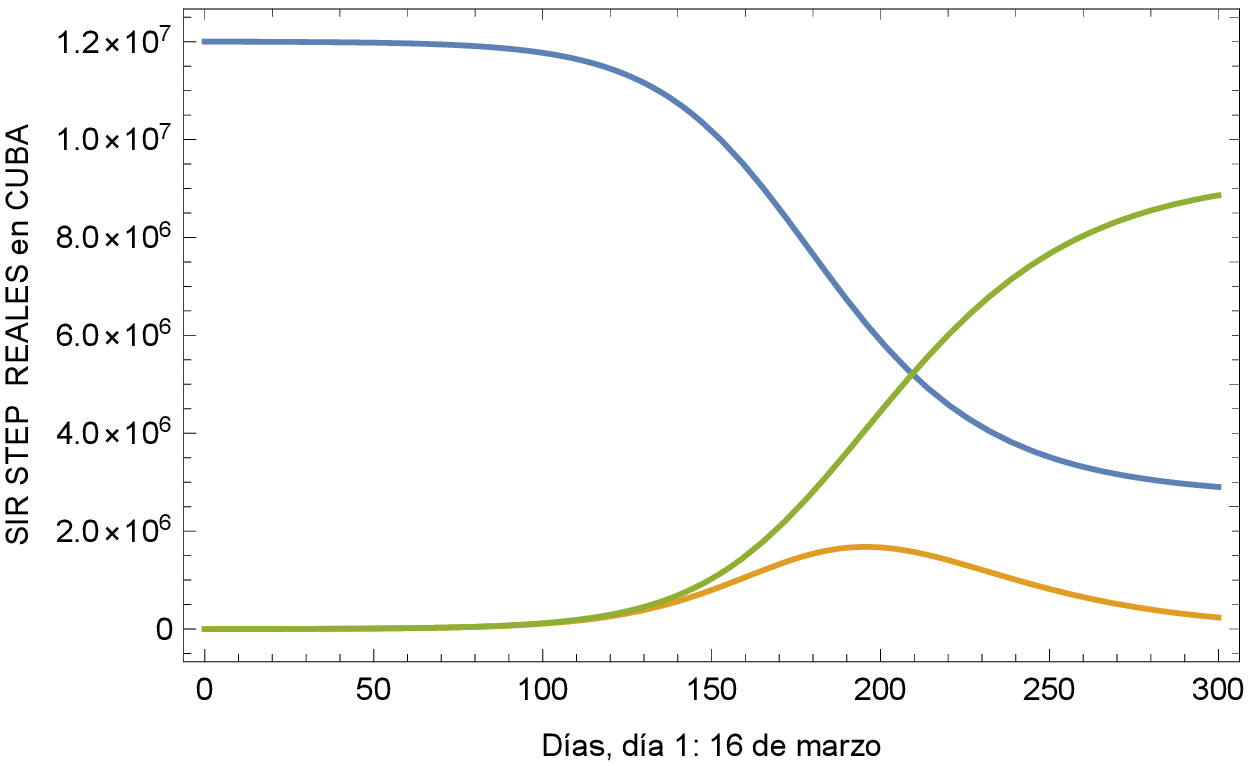}
\includegraphics[width=.45\textwidth]{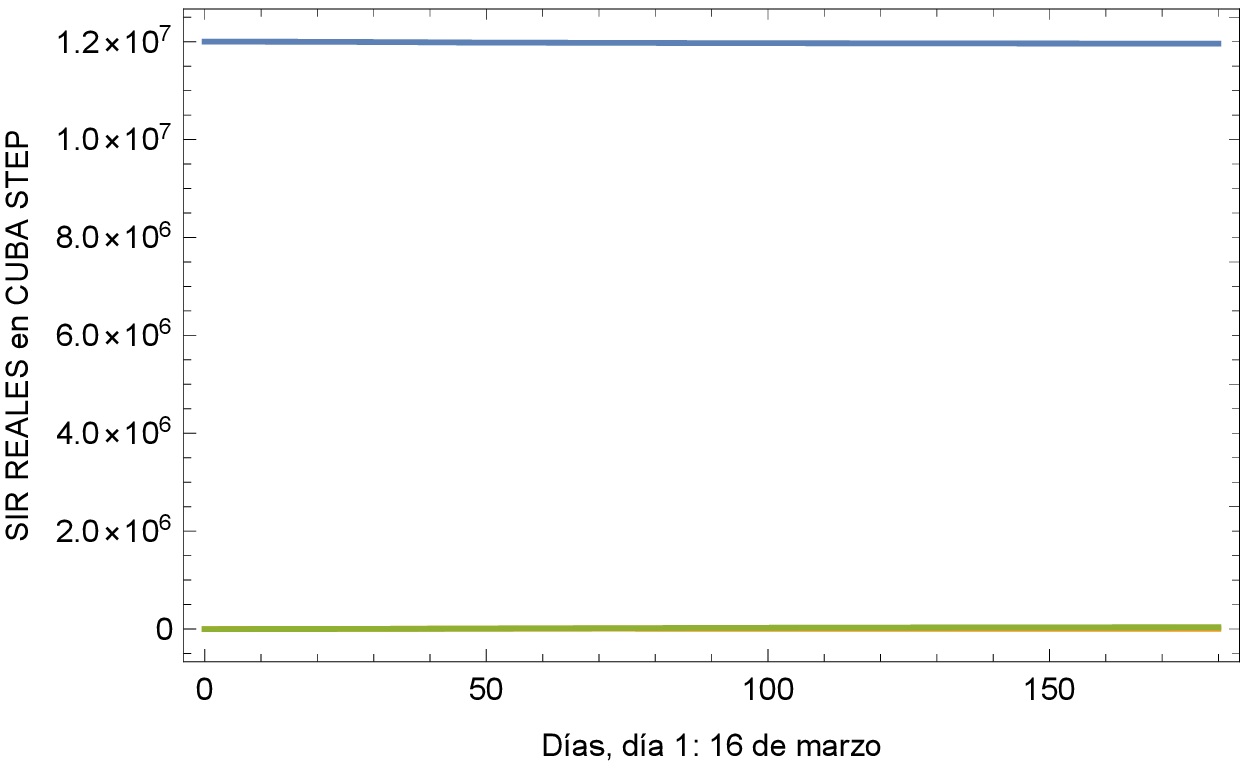}
\end{center}
\caption{{\protect\footnotesize Evoluci\'{o}n de suceptibles, infectados y
recuperados en el modelo SIR aplicado a Cuba. La curva azul representa los
susceptibles, la roja los recuperados y la amarilla los infectados. En el
panel de la izquierda  se ve que con la actual tasa de contagio se estima que
4 millones de personas no sufran contagio. El panel derecho  muestra co\'{m}o
ejemplo el efecto de un confinamiento estricto a finales de abril: una
epidemia controlada como parece ser el caso de China y Alemania, en que le
n\'umero de infectados es muy inferior a la poblacion.}}%
\label{ajusteE3}%
\end{figure}

 En la figura \ref{ajusteE} se muestra la
soluci\'{o}n del SIR, empleando los valores $\beta^{\ast}$ obtenidos haciendo
ajustes exponenciales locales mediante m\'{\i}nimos cuadrados. El panel
izquierdo de la figura \ref{ajusteE} muestra el ajuste a los datos, y el panel
derecho de  \ref{ajusteE} destaca que suceder\'{\i}a a la curva si a finales
de abril disminuye la tasa de contagio al nivel $\beta^{\ast}-\gamma<0$.
\begin{figure}[h]
\begin{center}
\includegraphics[width=.45\textwidth]{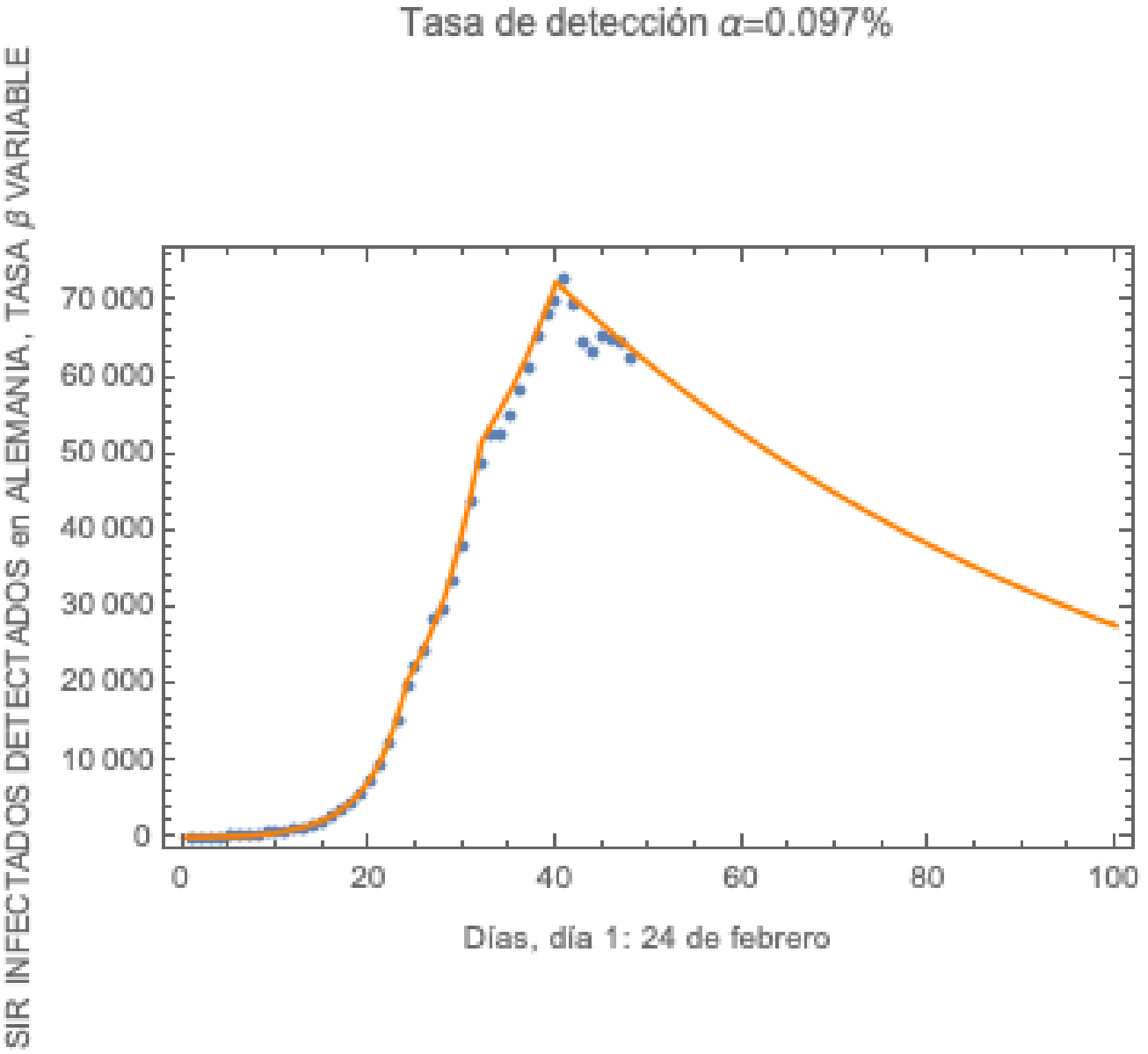}
\includegraphics[width=.45\textwidth]{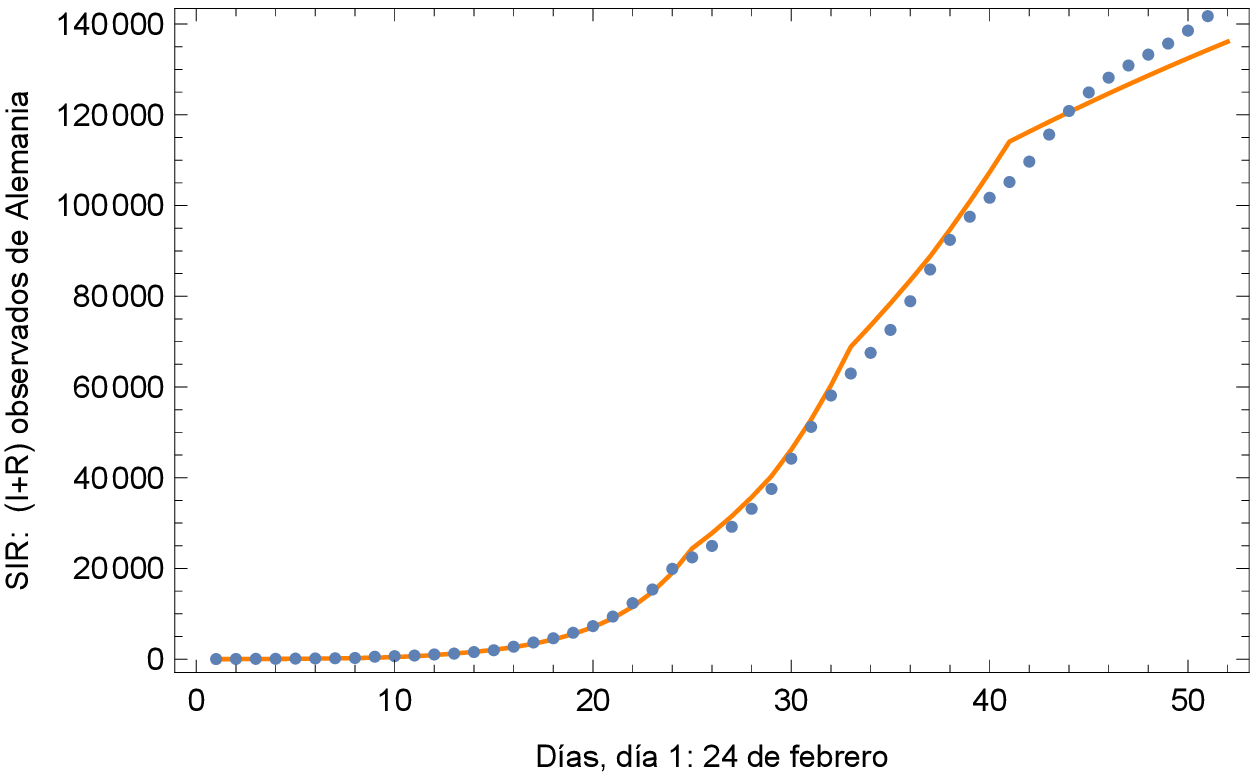}
\includegraphics[width=.45\textwidth]{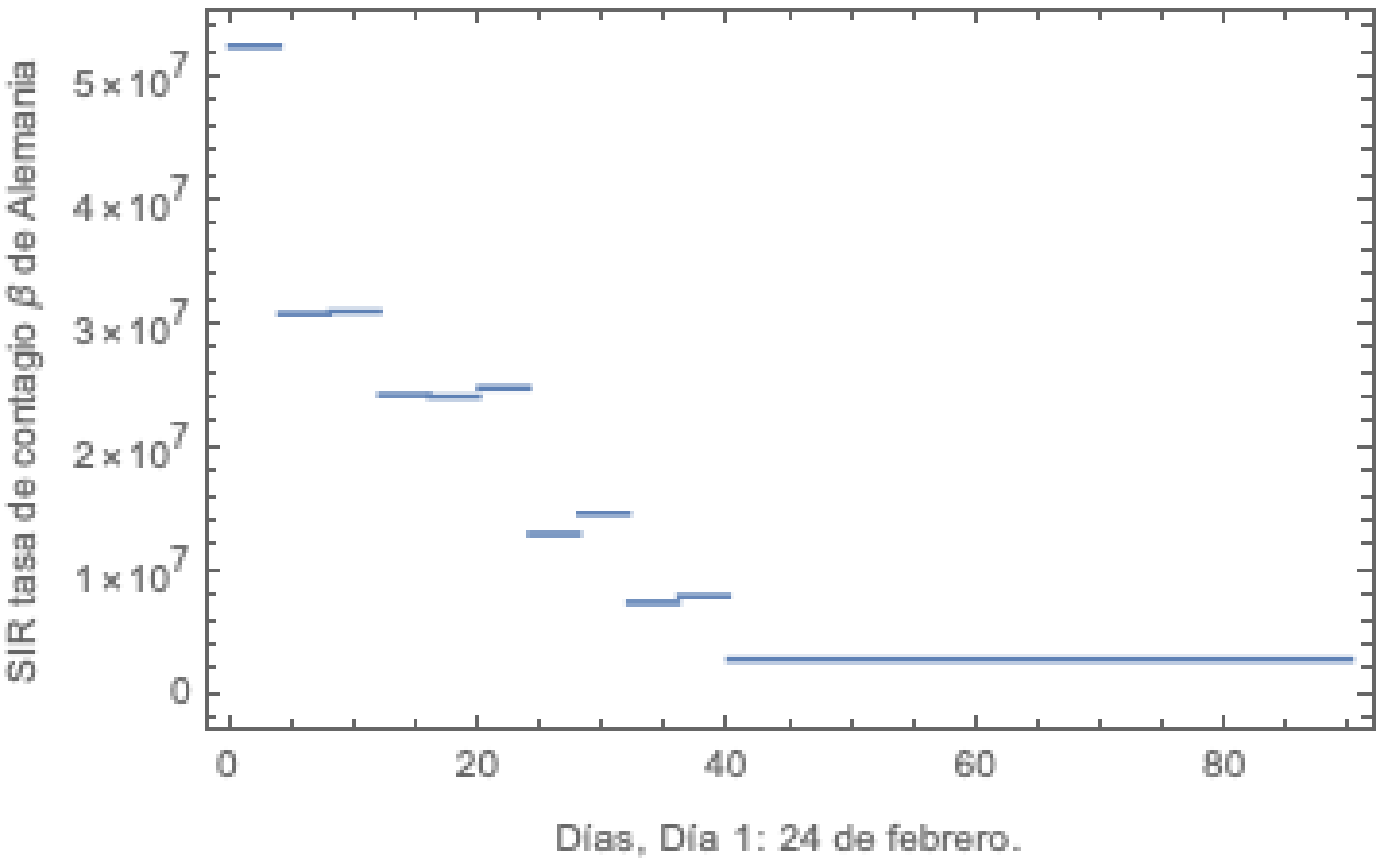}
\includegraphics[width=.45\textwidth]{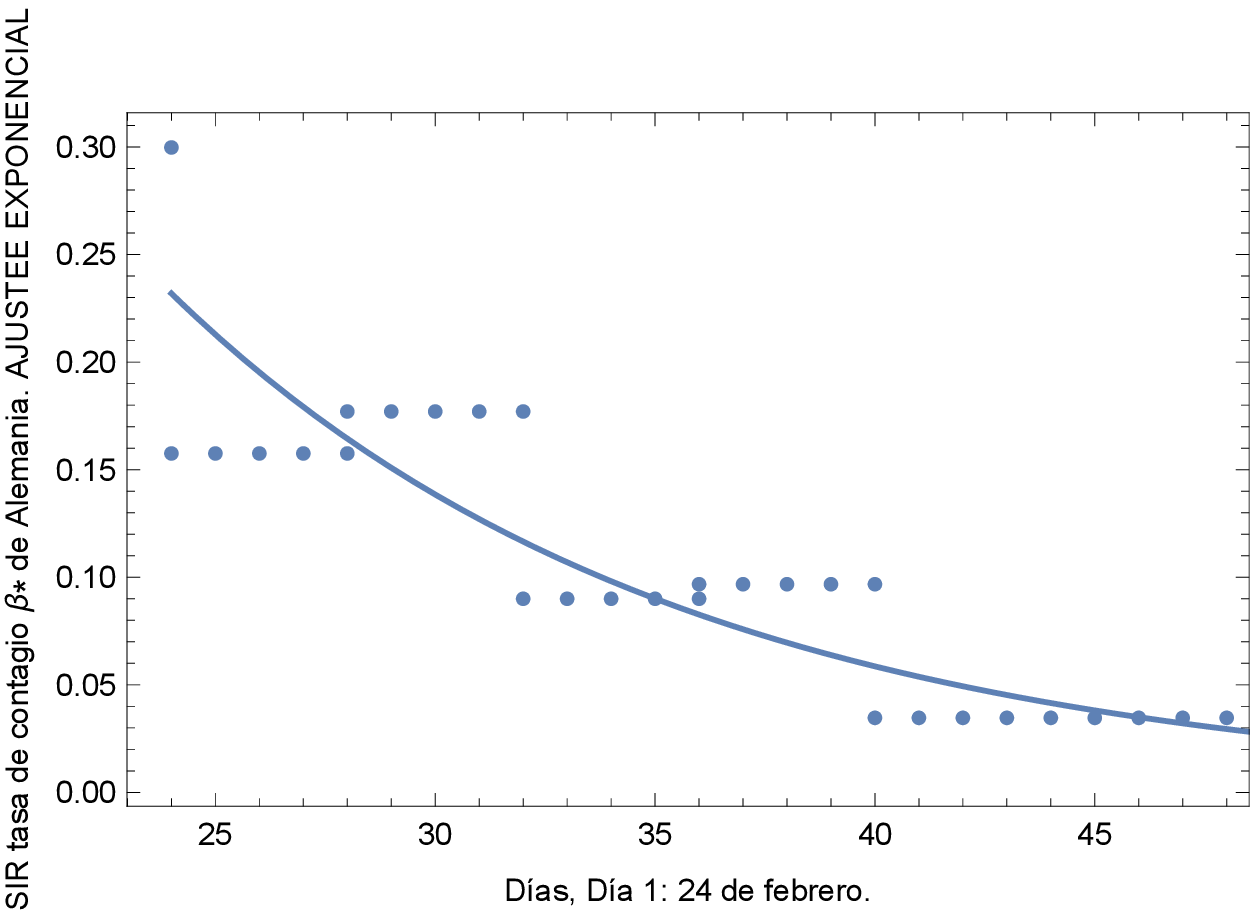}
\end{center}
\caption{{\protect\footnotesize En la primera fila el panel de la izquierda  muestra la
evoluci\'{o}n de los infectados para Alemania de acuerdo a los datos
experimentales (puntos azules) en a un modelo SIR con tasa de contagio $\beta$
variable (l\'{\i}nea naranja). En la primera fila el segundo gr\'afico muestra el n\'umero
de $I+R$ de los datos (puntos azules) confrontado con el modelo (l\'{\i}nea naranja). En la segunda fila se muestra primero la variaci\'{o}n de la tasa de contagio en el tiempo, y en el segundo gr\'afico se superpone el ajuste de una de exponencial (l\'{\i}nea) a las $\beta$s locales (puntos) a partir del confinamiento. Se considera $\gamma=1/20$. Alemania se encuentra en la regi\'{o}n
donde $\beta^{\ast}-\gamma<0$.}}%
\label{ajusteDE}%
\end{figure}

La figura \ref{ajusteE2} muestra en la primera columna la
situaci\'{o}n hipot\'{e}tica de evoluci\'{o}n del modelo SIR si la tasa de
contagio continua siendo la observada al \ 16.04.20. El pico de la
infecci\'{o}n ser\'{\i}a dilatado hasta septiembre, y alcanzar\'{\i}a los
200,000 infectados activos observados.
\begin{figure}[b]
\begin{center}
\includegraphics[width=.45\textwidth]{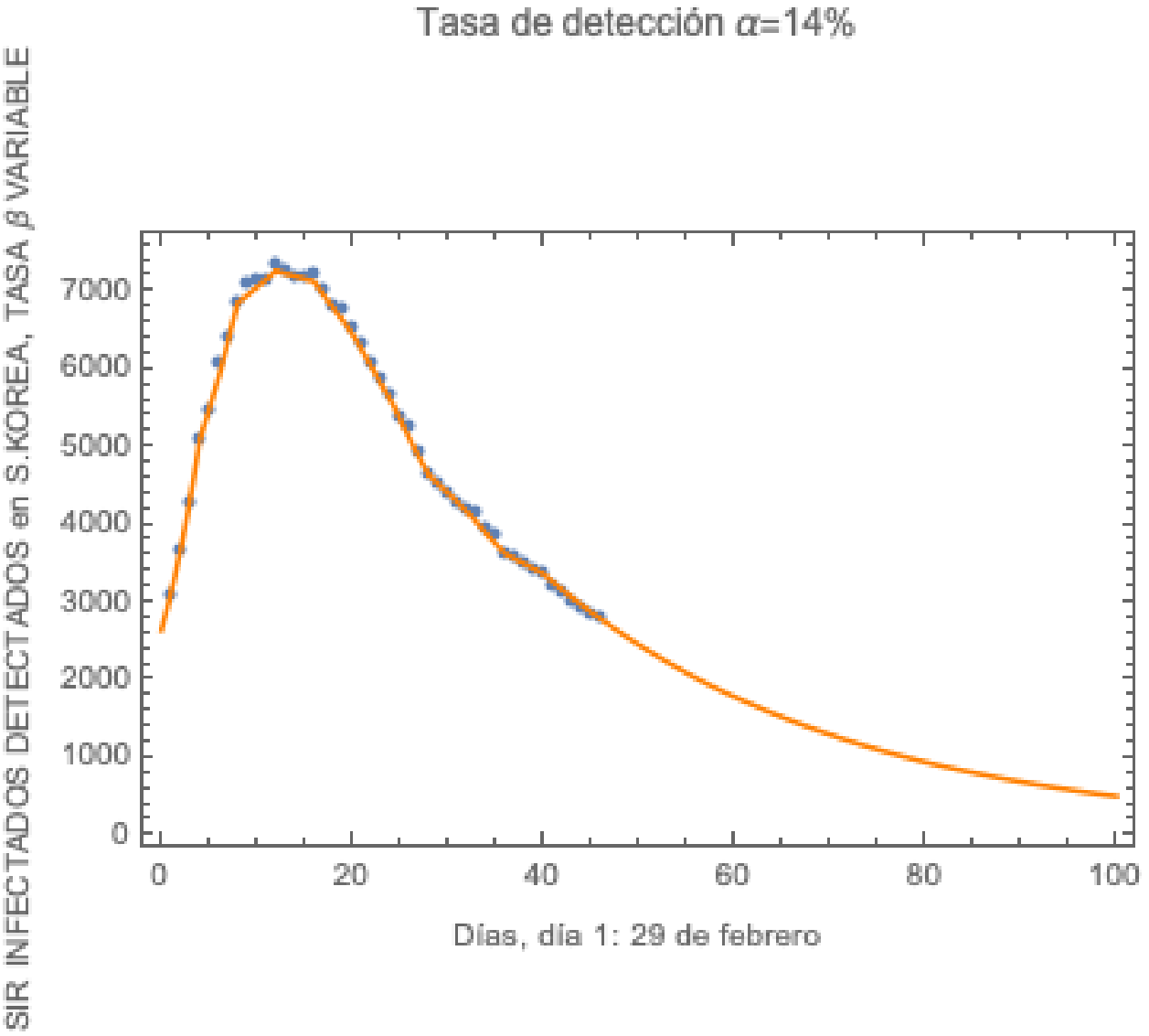}
\includegraphics[width=.45\textwidth]{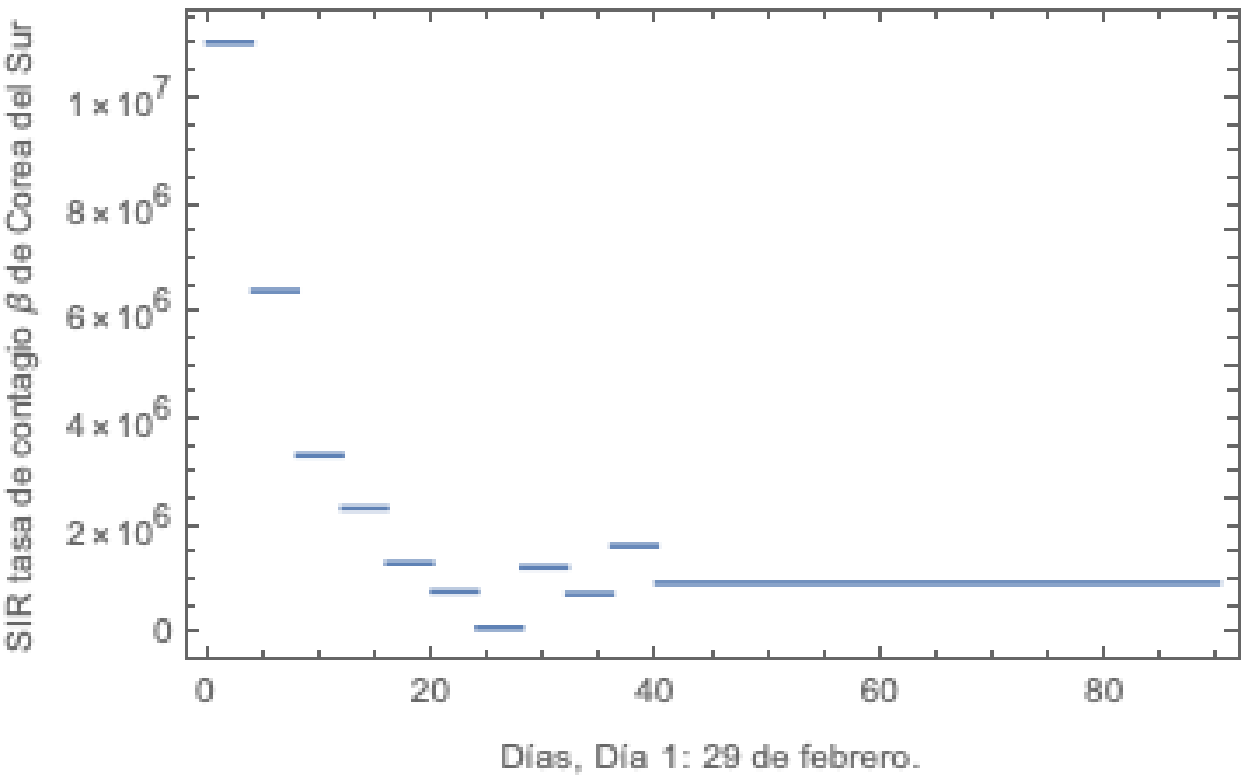}
\end{center}
\caption{{\protect\footnotesize El panel de la izquierda  muestra la
evoluci\'{o}n de los infectados para Corea del Sur de acuerdo a los datos
experimentales (puntos azules) y a un modelo SIR con tasa de contagio $\beta$
variable. El panel derecho  muestra la variaci\'{o}n de la tasa de contagio en
el tiempo. Se considera $\gamma=1/20$. Corea del Sur se encuentra en la
regi\'{o}n donde $\beta^{\ast}-\gamma<0$.}}%
\label{ajusteKOR}%
\end{figure}
Se considera una tasa de detecci\'{o}n
$k=0.14$ \cite{5,12} lo que dir\'{\i}a que el pico real seria de
aproximadamente 1.5 millones de personas. Este escenario dir\'{\i}a que la
mayor parte de la poblaci\'{o}n del pa\'{\i}s se enfermar\'{\i}a, como se
puede ver en el panel izquierdo de la  figura \ref{ajusteE3}, y no es deseable. De aqu\'{\i} la
necesidad de confinamiento.
En la segunda columna de la figura \ref{ajusteE2} se muestra la
situaci\'{o}n hipot\'{e}tica que debido a las medidas de confinamiento, a
finales de abril se alcance un $\beta^{\ast}=0.0384309$ dado en la quinta
columna de la tabla \ref{pedazos}.
\begin{table}[t]
\begin{center}%
\begin{tabular}
[c]{|c|c|c|c|}\hline
Per\'{\i}odos & $\gamma(\tilde R_{0}-1)$ & $\tilde\beta$ & $\tilde R_{0}%
$\\\hline
(1): 26.02.20 \text{ al } 29.02.20 & 0.583043 & $5.25552*10^{7}$ &
12.6609\\\hline
(2): 29.02.20 \text{ al } 04.03.20 & 0.32193 & $3.08777*10^{7}$ &
7.43861,\\\hline
(3): 04.03.20 \text{ al } 08.03.20 & 0.324113 & $3.10589*10^{7}$ &
7.48226\\\hline
(4): 08.03.20-\text{ al } 12.03.20 & 0.242933 & $2.43193*10^{7}$ &
5.85866\\\hline
(5): 12.03.20-\text{ al } 16.03.20 & 0.241112 & $2.41682*10^{7}$ &
5.82225\\\hline
(6): 16.03.20-\text{ al } 20.03.20 & 0.249774 & $2.48873*10^{7}$ &
5.99549\\\hline
(7): 20.03.20-\text{ al } 24.03.20 & 0.10755 & $1.30798*10^{7}$ &
3.15101\\\hline
(8): 24.03.20-\text{ al } 28.03.20 & 0.127047 & $1.46985*10^{7}$ &
3.54094\\\hline
(9): 28.03.20-\text{ al } 01.04.20 & 0.0400058 & $7.47228*10^{6}$ &
1.80012\\\hline
(10): 01.04.20-\text{ al } 05.04.20 & 0.04677 & $8.03384*10^{6}$ &
1.9354\\\hline
(11): 05.04.20-\text{ al } 14.04.20 & -0.0161831 & $2.80747*10^{6}$ &
0.676337\\\hline
\end{tabular}
\end{center}
\caption{Tasa de contagio variable a pedazos para Alemania. En el presente
per\'{\i}odo se alcanza $\tilde{R}_{0}<1$, por lo que la epidemia est\'{a} por
el momento controlada.}%
\label{2}%
\end{table}

\begin{table}[h]
\begin{center}%
\begin{tabular}
[c]{|c|c|c|c|}\hline
Per\'{\i}odos & $\gamma(\tilde R_{0}-1)$ & $\tilde\beta$ & $\tilde R_{0}%
$\\\hline
(1): 29.02.20 \text{ al } 03.03.20 & 0.164554 & $1.10431*10^{7}$ &
4.29108\\\hline
(2): 03.03.20 \text{ al } 07.03.20 & 0.0743729 & $6.40147*10^{6}$ &
2.48746\\\hline
(3): 07.03.20 \text{ al } 11.03.20 & 0.0146416 & $3.3271*10^{6}$ &
1.29283\\\hline
(4): 11.03.20-\text{ al } 15.03.20 & -0.00454485 & $2.33958*10^{6} $ &
0.909103\\\hline
(5): 15.03.20-\text{ al } 19.03.20 & -0.0244965 & $1.31267*10^{6}$ &
0.510071\\\hline
(6): 19.03.20-\text{ al } 23.03.20 & -0.0348858 & $777930.$ & 0.302285\\\hline
(7): 23.03.20-\text{ al } 27.03.20 & -0.0480769 & $98981.1$ &
0.0384617\\\hline
(8): 27.03.20-\text{ al } 31.03.20 & -0.0258793 & $1.24149*10^{6}$ &
0.482413\\\hline
(9): 31.03.20-\text{ al } 04.04.20 & -0.0357962 & $731069$ & 0.284076\\\hline
(10): 04.04.20-\text{ al } 08.04.20 & -0.0180901 & $1.64241*10^{6}$ &
0.638199\\\hline
(11): 08.04.20-\text{ al } 14.04.20 & -0.0317687 & $938363.$ &
0.364625\\\hline
\end{tabular}
\end{center}
\caption{Tasa de contagio variable a pedazos para Corea del Sur. En el
presente per\'{\i}odo se alcanza $\tilde{R}_{0}<1$, por lo que la epidemia
est\'{a} por el momento controlada.}%
\label{3}%
\end{table}
En este caso el pico de infectados activos se
alcanzar\'{\i}a a finales de mayo y los valores observados estar\'{\i}an en el
orden de los 1200, los reales m\'{a}s bien en los 8,000 casos. Esto se puede
ver tambi\'{e}n en el panel derecho de la  figura \ref{ajusteE3}  donde se
observa claramente que la poblaci\'{o}n infectada ser\'{\i}a una peque\~{n}a
fracci\'{o}n de los 12 millones de habitantes. Este escenario ideal es una
muestra clara de que el m\'{a}ximo que queremos alcanzar no es el m\'{a}ximo
standard de la evoluci\'{o}n del SIR, si no un m\'{a}ximo logrado cuando
$\tilde{R}_{0}<1$. \ Este es una propiedad importante a tener en cuenta por
las personas  que en estos tiempos deseen comenzar a analizar esta pandemia.

  Para aclarar a\'{u}n m\'{a}s nuestra idea mostraremos las curvas
de Alemania y Corea del Sur, esta ultima ya en fase de salida de la epidemia.
Las figuras \ref{ajusteDE} y \ref{ajusteKOR} muestran la
evoluci\'{o}n de la epidemia en estas dos naciones empleando un modelo SIR con
tasa de contagio variable en el tiempo, con vistas a reflejar el efecto del
confinamiento.
Las l\'{\i}neas s\'{o}lidas representan a las soluciones de las
ecuaciones diferenciales del modelo SIR empleando la tasa $\beta(t)$ obtenida
de realizar ajustes locales. En las Tablas \ref{2} y \ref{3} se representa la
evoluci\'{o}n de los valores de las tasas de contagio $\beta$, $\tilde{R}_{0}$
y $\gamma(\tilde{R}_{0}-1)$. Se observa que ambos pa\'{\i}ses alcanzaron la
regi\'{o}n $\tilde{R}_{0}<1$.

\newpage

\section{Conclusiones}

En el trabajo se exploraron dos descripciones de la pandemia. Una de ellas
constituye un modelo emp\'{\i}rico en el cual se considera que desde el punto
en que se establece el confinamiento, la tasa de contagio debe, al menos,
empezar a  decrecer tendiendo  cero  despu\'es  de un per\'{\i}odo de duraci\'{o}n de la
enfermedad ($1/\gamma$) (aproximadamente  15 a 20 d\'{\i}as).  Este modelo es
capaz de describir  bien el comportamiento de la epidemia en pa\'{\i}ses como
Alemania. El segundo es un modelo SIR con tasa de contagio variable, la
cual se ajusta experimentalmente, y su variaci\'{o}n se asume que se produce a
consecuencia de la cuarentena. Estos an\'alisis sugieren que la tasa de contagio a 
partir del d\'{\i}a de aislamiento se comporta aproximadamente como una exponencial. 
Ambas descripciones coinciden, y el enf\'{a}sis de ambos an\'{a}lisis radica en que un control de la epidemia s\'{o}lo se
logra con un confinamiento riguroso. \ Por lo cual analizando la evoluci\'{o}n
de la epidemia, resulta  muy importante el monitorear los valores de $\beta$
instant\'{a}neos para medir si el confinamiento est\'{a} siendo lo estricto
que se necesita. Esto es, con vistas a controlar la pandemia con un n\'umero de
infectados muy inferior a la poblacion se requiere  que en cierto per\'{\i}odo
de d\'{\i}as se cumpla la condicion $\beta^{\ast}-\gamma<0$ ($\tilde{R}_{0}%
<1$). En los d\'{\i}as del  12-04-20 al 18.04.20 la tasa de contagio de Cuba
est\'{a} en $\tilde{R}_{0}=1.92154$, \ por lo que a\'{u}n no han sido
totalmente efectivas las medidas de confinamiento (Alemania posee una tasa de
0.68 determinada en este trabajo y reportada tambi\'{e}n en la literatura).

El modelo para la evoluci\'{o}n de $\beta$ aplicado al caso cubano, predice
que la epidemia pudiera concluir inclusive a finales de abril o mediados de
mayo si las medidas de aislamiento tomadas en el entorno del 24 de Marzo,
resultaran efectivas. En este caso optimista, el m\'{a}ximo n\'{u}mero de
infectados podr\'{\i}a resultar en cerca de $1000-2000$ infectados activos con
el m\'{a}ximo apareciendo entre finales de abril y el 12 de mayo. El valor del
m\'{a}ximo crecer\'{\i}a a medida que este tarde mas en realizarse.

Tambi\'en en los modelos de $\beta$ variables a pedazo en el tiempo, se
asumi\'o una implantaci\'on efectiva  m\'as tard\'ia del aislamiento a finales
de abril, que  podr\'{\i}a definir a finales de mayo un pico de infectados
activos del orden de los miles de personas. En esta situaci\'on optimista el
n\'umero de personas m\'aximo en estado grave estar\'{\i}a en el orden los
cientos. En el caso de no lograr controlar la evoluci\'on de la epidemia, y
continuar con la tasa de contagio al d\'{\i}a de hoy los casos graves
rondar\'{\i}an los 20000, y este pico se alcanzar\'{\i}a en el mes de septiembre.

Nos gustar\'{\i}a destacar que el mensaje importante de este trabajo es el
hecho de que la din\'{a}mica de los par\'{a}metros que controlan la
evoluci\'{o}n de la epidemia es relevante. En particular la evoluci\'{o}n de
la tasa de contagio $\beta$ (o de  $\tilde{R}_{0}-1,$ o de $\frac{I^{\prime
}(t)}{I(t)}$) debe tomarse muy en cuenta  para determinar que medidas
imponer. La observaci\'on de estas cantidades sin duda permite  una estimaci\'on
cualitativa de cuando se  debe obtener el m\'aximo de casos, es decir, de cuando comenzar\'a
el fin de la epidemia. \

\section{Agradecimientos}

Agradecemos a Juan Barranco, Argelia Bernal, Alejandro Cabo Bizet, Milagros Bizet, David Delepine, Alejandro Gil, Augusto Gonz\'alez, Alma Gonz\'alez, Alejandro Lage, Oscar Loaiza, Albrecht Klemm, Dami\'an Mayorga, Mauro Napsuciale, Gustavo Niz, Octavio Obreg\'on y Luis Ure\~{n}a, por discusiones muy \'utiles y comentarios durante el desarrollo del trabajo. NCB
agradece al Proyecto CONACyT A1-S- 37752 y al Laboratorio de Datos, DCI, de la
Universidad de Guanajuato.  ACM desea agradecer el apoyo recibido del Network 09,
de la Oficina de Asuntos Externos (OEA) del ``Centro Internacional para la F\'isica Te\'orica"
(ICTP) en Trieste, Italia.

\end{document}